\documentclass[twocolumn,prb,raggedbottom,floatfix]{revtex4}
\usepackage{amsmath}
\usepackage{graphicx} 
\usepackage{color} 
\usepackage{textcomp}
\usepackage{url}
\usepackage[none]{hyphenat}
\usepackage{placeins}

\begin{document}

\title{Stability and superconductivity of Ca-B phases at ambient and high pressure}

\author{Sheena Shah$^1$ and Aleksey N. Kolmogorov$^2$}
\affiliation{$^1$University of Oxford, Department of Materials, Parks Road, Oxford OX1 3PH, United Kingdom}
\affiliation{\mbox{$^2$Department of Physics, Applied Physics and Astronomy, Binghamton University, State University of New York,} PO Box 6000, Binghamton, New York 13902-6000}
\date{\today}

\begin{abstract}
{In the search for MgB$_2$-like phonon-mediated superconductors we have carried out a systematic density functional theory study of the Ca-B system, isoelectronic to Mg-B, at ambient and gigapascal pressures.  A remarkable variety of candidate high-pressure ground states have been identified with an evolutionary crystal structure search, including a stable alkaline-earth monoboride oI8-CaB, a superconductor with an expected critical temperature (\textit{T$_c$}) of 5.5 K.  We have extended our previous study of CaB$_6$ [Phys. Rev. Lett. {\bf 108}, 102501 (2012)] to nearby stoichiometries of CaB$_{6+x}$, finding that extra boron further stabilizes the proposed B$_{24}$ units.  Here an explanation is given for the transformation of cP7-CaB$_6$ into the more complex oS56 and tI56 polymorphs at high pressure.  The stability of the known metallic tP20 phase of CaB$_4$ at ambient pressure is explained from a crystal structure and chemical bonding point of view.  The tP20 structure is shown to destabilize at 19 GPa relative to a semiconducting MgB$_4$-like structure due to chemical pressure from the metal ion.  The hypothetical AlB$_2$-type structure of CaB$_2$, previously shown to have favorable superconducting features, is demonstrated here to be unstable at all pressures; two new metallic CaB$_2$ polymorphs with unusual boron networks stabilize at elevated pressures above 8 GPa but are found to have very low critical temperatures (\textit{T$_c$} $\sim$ 1 K).  The stability of all structures has been rationalized through comparison with alkaline-earth analogs, emphasizing the importance of the size of the metal ion for the stability of borides.  Our study illustrates the inverse correlation between the thermodynamic stability and superconducting properties and the necessity to carefully examine both in the design of new synthesizable superconducting materials.}

\end{abstract}

\maketitle

\section{Introduction}
The development of computational crystal structure prediction tools \cite{Curtarolo2013, Hautier2012} and methods to calculate electron-phonon based superconducting properties \cite{Eliashberg1960, Allen1975, Margine2013, CohenBCS} have led to a large number of BCS superconductors being \mbox{theoretically} proposed.  Experimentally confirmed predictions, such as the superconductivity of already known structures of silicon \cite{Chang1984} and lithium \cite{Neaton1999, Shimizu2002} under pressure, are however rare as designing superconductors requires careful management of two, often competing, requirements.  First, one must make sure that predicted metallic crystal structure phases are thermodynamically stable with respect to decomposition/transformation into more stable phases or at least dynamically stable with respect to all possible distortions.  Second, one typically screens viable structures for signature electronic and vibrational characteristics observed in known BCS superconductors with relatively high critical temperatures (\textit{T$_c$}) \cite{CohenBCS}.  Features thought to be favorable include a high electronic density of states (DOS) at the Fermi level (\textit{E$_F$}) \cite{Pickett2008} and a large softening of certain high-frequency phonon modes. \cite{Buzea2005, Lortz2006}  Materials composed of elements with a low mass or materials which are close to structural instability have been considered as promising candidates to possess such properties and therefore could make good BCS superconductors.  

The difficulty comes from combining stability with superconducting properties as shown in recent \cite{Flores-Livas2012, Kolmogorov2008} and the present studies.  For example, one could say that the superconducting properties of the analog of MgB$_2$, \cite{An2001} CaB$_2$, are ideal.  It is composed of light elements and, relative to MgB$_2$, it has a higher DOS at the Fermi level and a more pronounced vibrational phonon softening of the in-plane boron mode. \cite{Kolmogorov2008}  A recent study on the \mbox{hypothetical} hP3-CaB$_2$ (Pearson notation) structure indeed predicted it to be a superconductor with a \textit{T$_c$} of 50 K. \cite{Choi2009}  Although dynamically stable, the structure is thermodynamically unstable relative to elemental \mbox{calcium} and \mbox{calcium} tetraboride \cite{Kolmogorov2006} as well as to our proposed polymorph oI12 based on four- and eight-membered ring \mbox{layers} of boron.

Various other MgB$_2$ type superconductors have been proposed, such as Li$_x$BC, \cite{Rosner2002} LiB \cite{Calandra2007} and Li$_2$AlB$_4$ \cite{Kolmogorov2008} but \mbox{experimental} observations of their superconducting properties are still unconfirmed. \cite{Lazicki2010,Bharathi2002}  To extend the search for MgB$_2$ type superconductors, here we have studied the whole composition range of Ca$_x$B$_{1-x}$, isoelectronic to Mg-B, at both ambient and high pressure.  Calcium under pressure currently has the highest \textit{T$_c$} amongst the elements of {\color{black} 29 K. \cite{Fujihisa2013}}  In addition, elemental calcium{\color{black}\cite{Oganov2010,Fujihisa2013}} and boron \cite{Oganov2009,Zarechnaya2008} have been shown to form complex high-pressure structures giving us further motivation to study their compounds.
\begin{figure*}[t]
\includegraphics[width = 16cm]{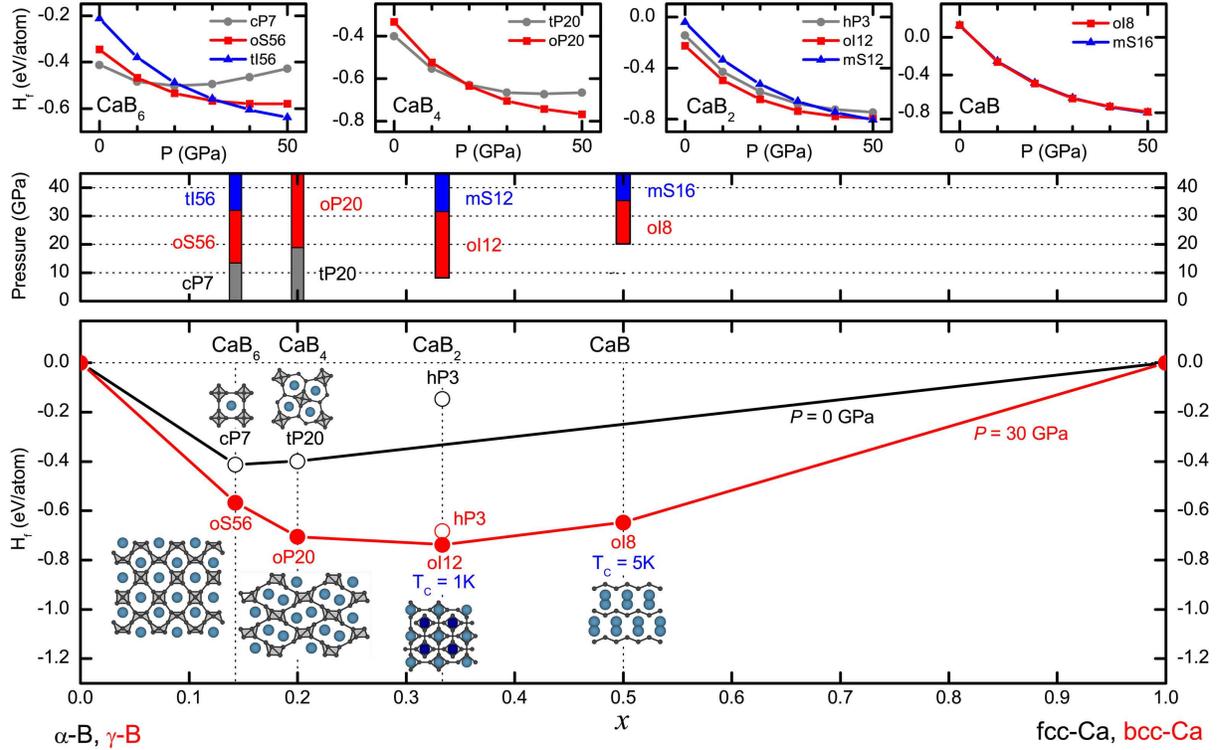}
\caption{(Color online) Top: the formation enthalpy of proposed structures of CaB$_6$, CaB$_4$, CaB$_2$ and CaB.  The plot for CaB$_2$ also shows the unstable hP3 structure in gray.  mS16-CaB cannot be seen as its enthalpy is close to oI8-CaB.  Middle: the pressure ranges in which the proposed ground state structures are stable.  Bottom: calculated formation enthalpies of Ca$_x$B$_{1-x}$ compounds at 0 (black) and 30 GPa (red) with their structures and superconducting critical temperatures (\textit{T$_c$}).  }
\label{fig1}
\vspace{-0.1cm}
\end{figure*}

In the set of Ca-B crystal structures we have considered, medium concentrations of calcium (0.33-0.50) are found to form stable superconducting compounds.  The superconductors all have a high contribution of \mbox{calcium} electronic states at the Fermi level.  Calcium rich \mbox{structures} ($x$ $>$ 0.50) are found to be unstable.  An \mbox{evolutionary} ground state crystal structure search has identified four new high-pressure superconducting structures: oI8-CaB (\textit{T$_c$} = 5.5 K),  oI12-CaB$_2$ (\textit{T$_c$} $\sim$ 1 K), mS12-CaB$_2$ (\textit{T$_c$} $\sim$ 1 K) and CaB$_{6.25}$ (\textit{T$_c$} $\approx$ 1.9 K).  We discuss the behavior of the stability with pressure of \mbox{calcium} and other alkaline-earth metal borides in the proposed ground state structures.  
 
{\color{black} Our study of CaB$_4$ agrees with previous DFT-ba sed calculations showing that tP20-CaB$_4$ should be stable at ambient pressure. \cite{Chepulskii2009}  However there have been concerns about its stability when considering the electron count within the compound. \cite{Yahia2008}  Here we explain why the ThB$_4$-type CaB$_4$ structure should be stable} at 0 GPa but destabilizes at 19 GPa relative to a semiconducting MgB$_4$ type structure resulting from out-of-plane chemical pressure from the metal ion.  Expanding upon our past investigation of CaB$_6$, \cite{Kolmogorov2012} we show that the phase transition to oS56 at 13 GPa is correlated to the length of one specific boron-boron bond length whilst the transition to tI56 at 32 GPa is due to space constraints for accommodating the metal ion.  Following the experiments supporting the tI56 parent structure, \cite{Kolmogorov2012} we find a more stable derivative, tP57-CaB$_{6.125}$, created by the insertion of an extra boron atom into half of the B$_{24}$ units.  The stability and superconducting \textit{T$_c$} of these structures are shown to be highly sensitive to the boron concentration.  At the \textit{M}B$_2$ composition we find that CaB$_2$ stabilizes at 8 GPa as a low-\textit{T$_c$} superconducting layered structure with four- and eight-membered boron rings (oI12), like those observed in the composite boron-carbon layers of CaB$_2$C$_2$. \cite{Albert1999}  A monoclinic structure (mS12), composed of a buckled sheet of boron hexagons, stabilizes above 32 GPa.  Finally, we propose a stable alkaline-earth monoboride (oI8-CaB) at 20 GPa composed of zig-zagged boron chains; the highest \textit{T$_c$} superconductor (5.5 K) found so far in the calcium borides.   
\raggedbottom

\section{Overview}
Figure \ref{fig1} gives a summary of predicted structures (details in Supplemental Material \cite{supmat}), the pressures they are stable over and the calculated \textit{T} = 0 K formation enthalpies.  This paper is split up into seven sections.  Section \ref{sec:method} details the computational methodology and the evolutionary crystal structure search.  A summary of the stability and properties of each Ca:B composition will be given in Secs. \ref{sec:CaB6}-\ref{sec:CaB}: \ref{sec:CaB6} - CaB$_{6+x}$; \ref{sec:CaB4} - CaB$_4$; \ref{sec:CaB2} - CaB$_2$ and \ref{sec:CaB} - CaB.  For proposed structures we have rationalized their stability and examined their electronic (DOS and band structures) and vibrational properties.  For identified metallic structures, we have assessed the strength of their electron-phonon coupling, by calculating the Eliashberg spectral function, \textit{$\alpha^{2}$F($\omega$)}, and estimated a superconducting transition temperature (\textit{T$_c$}) using the Allen-Dynes \cite{Allen1975} equation.  

\section{Computational details} \label{sec:method}
We used density functional theory (DFT) calculations to assess the thermodynamic stability of over a hundred Ca$_x$B$_{1-x}$ structures.  Some of these structures were taken from known metal boride and related configurations \cite{Kolmogorov2010} listed in the Inorganic Crystal Structure Database (ICSD). \cite{ICSD}  The rest of the structures were generated by performing an unconstrained structural optimization using an evolutionary algorithm within the Module for Ab Initio Structure Evolution ({\small MAISE}) \cite{maise}.  {\small MAISE} relies on {\small VASP} \cite{Kresse1996} to calculate the enthalpy of crystal structures and uses this information to perform a global optimization to find the lowest enthalpy ordered phase at a given elemental composition. \cite{Blochl1994, Pack1976, Pack1977, Perdew1996}  It enables full structural optimization starting with random or known configurations by passing on beneficial traits from parents to offspring via crossover and mutation operations. The settings for the evolutionary algorithm search were similar to those used in our previous study. \cite{Kolmogorov2010}

\begin{center}
\begin{table}[!b]{\small
\hfill{}
\begin{tabular}{l|c|c}\hline\hline
Compound	&	Unit cells Ca:B	&	H$_{rel}$ (meV/atom)	\\\hline
\hspace{0.3cm}CaB$_{12}$	&	1:12	             &	32	\\
\hspace{0.3cm}CaB$_7$	&	2:14	                 &	60	\\
\hspace{0.3cm}CaB$_6$	&	1:6, 2:12, 3:18, 4:24  &	0\\
\hspace{0.3cm}CaB$_5$	&	1:5, 2:10	             &	81	\\
\hspace{0.3cm}CaB$_4$	&	1:14, 2:8, 3:16, 4:16  &	0	\\
\hspace{0.3cm}CaB$_3$	&	2:6, 3:9, 4:12	       &	114	\\
\hspace{0.3cm}CaB$_2$	&	3:6, 4:8, 6:12	       &	0	\\
\hspace{0.3cm}Ca$_2$B$_3$	&	2:3, 6:9	         &	136	\\
\hspace{0.3cm}Ca$_5$B$_6$	&	5:6	               &	153	\\
\hspace{0.3cm}CaB	&	3:3, 4:4, 5:5, 6:6, 8:8	   &	0	\\
\hspace{0.3cm}Ca$_2$B	&	4:2, 8:4, 12:6	       &	44	\\ \hline\hline
\end{tabular}}
\hfill{}
\caption{List of unit cells used for the evolutionary search across the Ca$_x$B$_{1-x}$ composition range and the relative enthalpy (as described in the text) at that composition at 30 GPa.}
\label{tb:evolutionary_unitcells}
\end{table}
\end{center}

\vspace{-1cm}
Table \ref{tb:evolutionary_unitcells} lists the unit cells used in the evolutionary search at each stoichiometry and a relative enthalpy ({\textit H$_{rel}$}).  {\textit H$_{rel}$} is defined as the ordinate distance \mbox{between} the point corresponding to the most stable structure at a given composition and the line connecting two nearest ground states in the formation enthalpy plot (Fig. \ref{fig1}).  Gibbs energies were calculated using {\small PHON} \cite{Alfe2009} and it was found that the vibrational entropic contributions at elevated temperatures for structures at different compositions typically did not change by more than 10-15 meV/atom.  With no available experimental information we have focused on the most likely to occur compositions. \cite{Rodgers1993}  All proposed ground states have been tested for thermodynamic stability relative to other Ca$_x$B$_{1-x}$ compositions.  In addition, ground states have been tested for dynamic stability by calculating their phonon dispersions and checking for the absence of phonon modes with imaginary frequencies.  Phonon dispersion and electron-phonon coupling calculations were performed with the Quantum {\small ESPRESSO} package \cite{Giannozzi2009} using ultrasoft pseudopotentials \cite{kq_meshes}.  Dense \textit{k}- and \textit {q}-meshes \cite{supmat} ensured that phonon frequencies converged with respect to the \textit{k}- and \textit{q}-mesh to within 10 cm$^{-1}$.  Energy cutoffs of 40 and 320 Ry were used for the wave functions and charge density respectively. 

For superconducting properties, calculation of the Eliashberg function determined the logarithmic average phonon frequency $\left\langle\omega_{ln}\right\rangle$, and the strength of the electron phonon coupling $\lambda$, which were both entered into the Allen-Dynes equation [Eqn. \ref{eq:Allen-Dynes}] \cite{Allen1975} to estimate the superconducting \textit{T$_c$}.   The Coulomb pseudopotential $\mu^{*}$ was assumed to be a constant of 0.12 as this is typically a good estimate for these types of superconducting compounds. \cite{Savrasov1996}

\begin{equation} 
T_{c} = \frac{\omega_{ln}}{1.2} exp \left\{ \frac{-\left[1.04\left(1 + \lambda \right)\right]}{\lambda - \mu^{*}\left(1+ 0.62 \lambda \right)}\right\} 
\label{eq:Allen-Dynes}
\end{equation}

Crystal ionic radii \cite{Shannon1976} were used for stability analysis and further details about this choice are given in the \mbox{Supplemental} Material. \cite{supmat}  In addition, all crystal structure diagrams have been adapted from those generated using {\small VESTA} \cite{Momma2011}.  

\section{C\lowercase{a}B$_{6+x}$ (\lowercase{$x$} = 0, 0.25, 0.125)}  \label{sec:CaB6}
A large number of hexaborides (lanthanides, actinides and alkaline-earths) adopt the same simple cubic structure (cP7-CaB$_6$) of B$_6$ octahedra connected in a three-dimensional network with metal ions in between.  Hexaborides are of interest for their valence-fluctuation (SmB$_6$) and Kondo effects (CeB$_6$), \cite{Ji2011} superconducting properties (YB$_6$, \textit{T$_c$} = 7 K) and use as thermionic \mbox{emitters} (LaB$_6$). \cite{Ji2011}  CaB$_6$ in particular was thought to be weakly ferromagnetic \cite{Young1999} but the observation is now attributed to impurities in the sample. \cite{Rhyee2004}   

Twenty electrons per boron octahedron (B$_6$$^{2-}$) are required for the stability of cP7-\textit{ M}B$_6$. \cite{Longuet1954}  This criterion may not apply for CaB$_6$ at high pressure as X-ray diffraction (XRD) measurements have indicated a cubic to \mbox{orthorhombic} phase transition around 12 GPa \cite{Li2011a} (the observation was not confirmed in follow up experiments\cite{Li2011b, Kolmogorov2012}).  A theoretical high pressure study by Wei \textit{et al.} \cite{Wei2011} compared two structures: cP7 and oP14 (Pban), \cite{pban_LaB6} concluding that cP7-CaB$_6$ does not undergo a phase transition up to 100 GPa.  Performing a fully unconstrained search at a particular chemical composition \mbox{enabled} us to propose much lower enthalpy orthorhombic and tetragonal high pressure crystal structures. \cite{Kolmogorov2012}

\begin{figure}[b]
\begin{center}
\includegraphics[width=85mm]{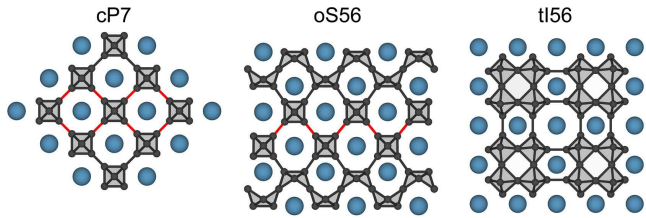}
\caption{(Color online) Structures of cP7, oS56 and tI56 CaB$_6$.  As for all structure diagrams in this article the boron atoms are in dark gray and calcium atoms in blue.  The bond between octahedra described in the text and in Fig. \ref{cP7_oS56_BB} is shown in red.}
\label{B04_oS56_8}
\end{center}
\vspace{-0.3cm}
\end{figure}

Using the DFT we predicted that the cP7 structure becomes dynamically unstable at 23 GPa \cite{Kolmogorov2012}.  At 13 GPa, cP7 is thermodynamically unstable relative to an orthorhombic oS56 structure consisting of B$_6$ octahedra and a chain of fused octahedra which open up and now share a bond (see Fig. \ref{B04_oS56_8}).  Upon applying further pressure (32 GPa) a tetragonal tI56 structure, made up of 24-atom units of fused boron octahedra, becomes thermodynamically stable. \cite{Kolmogorov2012}  The existence of the parent high pressure tI56 structure has been supported with powder XRD measurements \cite{Kolmogorov2012}. 

High pressure phase transitions have been observed in other metal hexaborides such as LaB$_6$ and YB$_6$.  The LaB$_6$ structure was indexed as Pban \cite{Teredesai2004} but was not observed in subsequent experiments. \cite{Godwal2009}  Our phonon calculations also indicate no reason for the cP7 structure to distort up to 60 GPa with the proposed Pban structure relaxing back to cP7 at 15 GPa.  YB$_6$ is superconducting with a \textit{T$_c$} of 7.1 K which has been predicted to drop with pressure.  The \textit{R} and \textit{M} phonon modes of YB$_6$ become imaginary at high pressures (55 and 63 GPa, respectively) \cite{Xu2007} like in CaB$_6$ \cite{Kolmogorov2012}.  

At ambient pressure, MgB$_6$ has been suggested to stabilize as a related cubic structure which exhibits surprising magnetic and antiferroelectric properties. \cite{Popov2012}  In order to generate the proposed stable structure, only one displacement from cP7 was considered. In this displacement, the magnesium ion sits closer to the face of the unit cell that had been artificially forced to remain cubic in the simulation.  Our analysis shows that the structure is unstable to decomposition into experimentally known MgB$_7$ and MgB$_4$ by 95 meV/atom.  We have also found more stable configurations for MgB$_6$: the proposed oS56 ground state of CaB$_6$ and local minima around cP7-CaB$_6$ of oP28, hR42 and tI28.  The lowest energy configuration, oS56, is 37 meV/atom lower in enthalpy than the proposed magnetic cubic structure but still 59 meV/atom above the MgB$_7$ $\leftrightarrow$ MgB$_4$ tie-line.              

\subsection{Structural properties of CaB$_6$ and analysis of the dynamical instability}\label{sec:cab6_structure}

\begin{figure}[h!]
\begin{center}
\includegraphics[width=85mm]{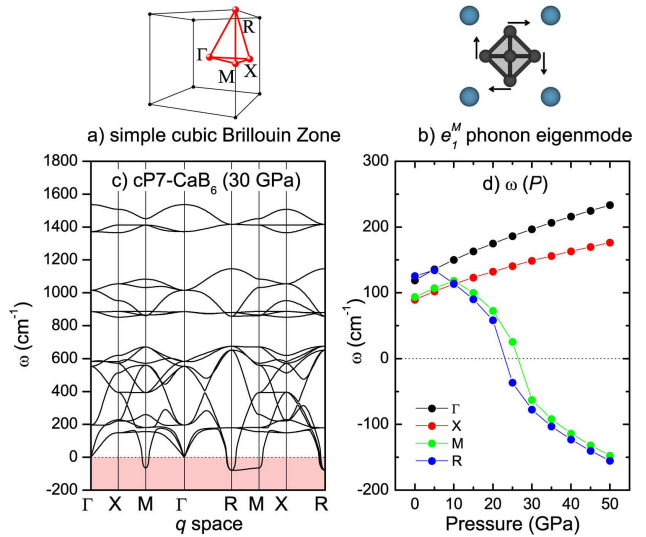}
\caption{(Color online) (a) High symmetry points defining the irreducible wedge in the simple cubic Brillouin zone. (b) Displacements of boron positions in real space viewed along [100] corresponding to the $M$-point phonon. (c) Calculated phonon dispersion for cP7-CaB$_6$ at 30 GPa showing multiple imaginary modes. (d) Calculated variation of $R$ and $M$ phonon frequencies with pressure.}
\label{fig6}
\end{center}
\end{figure}

For dynamical stability analysis we calculated the phonon dispersion of cP7-CaB$_6$ under pressure (Fig. \ref{fig6}) and found a whole branch along the $M$-$R$ direction becoming imaginary above 23 GPa (see Ref. \cite{supmat}).  Frequencies of selected phonon modes at the $\Gamma$, $X$, $M$, and $R$ high-symmetry points were calculated as a function of pressure [see Fig. \ref{fig6}(d)].  

To find the lower energy phase of CaB$_6$ we constructed possible distorted structures for the $2\times 2\times 2$ cP7 supercell by considering all non-equivalent combinations of the three degenerate $R$-point and one $M$-point eigenvectors as listed in Table \ref{table_cab6_eigen}.  The resulting distorted unit cells were carefully relaxed (the residual forces were below 0.001 eV/atom) and found to converge to several distinct configurations judging by their space group, enthalpy, atomic volume, simulated XRD pattern \cite{supmat} and radial distribution functions (see Ref. \cite{Kolmogorov2012}).  For example, the most general distortion $e^R_1+e^R_2+e^R_3+e^M_1$ led to the same oP28 structure as the $e^R_2+e^R_3+e^M_1$ distortion.  For each distinct new structure we ran a phonon dispersion calculation and established that the oP28, hR42 and tI28 structures \cite{supmat} of CaB$_6$ are the local minima around the parent cP7 structure. 

\begin{figure}[t!]
\begin{center}
\includegraphics[width=70mm]{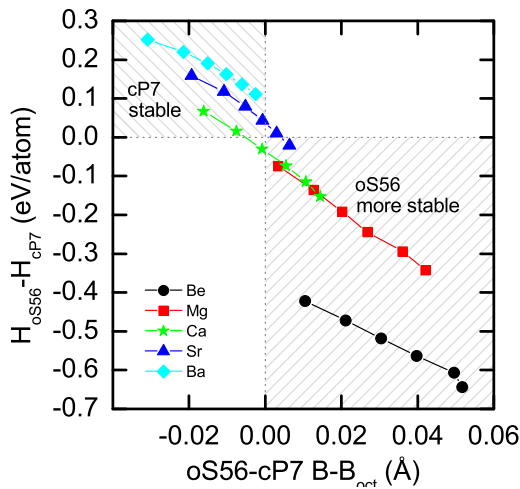}
\caption{(Color online) Difference between boron-boron bond length between octahedra (B-B$_{oct}$) in oS56 and cP7 against the difference in their formation enthalpies for all alkaline-earth metals at increasing pressures from left to right of 0, 10, 20, 30, 40 and 50 GPa.  The B-B bond is shown in red in Fig. \ref{B04_oS56_8}.  The kink at 50 GPa for the Be point is thought to be due to collapse of the highly unstable BeB$_6$. }
\label{cP7_oS56_BB}
\end{center}
\end{figure}

The evolutionary search at 30 GPa revealed the proposed ground state of oS56-CaB$_6$ which stabilizes over cP7 at 13 GPa.  Applying pressure to cP7-CaB$_6$ reduces the B-B bond between octahedra (from 1.673 {\AA} at 0 GPa to 1.571 {\AA} at 30 GPa).  The shortened bond breaks and half of the octahedra open up and fuse which results in the formation of zigzag strips in one direction (oS56 in Fig. \ref{B04_oS56_8}).  For all alkaline-earth hexaborides at 0 - 50 GPa, oS56 is found to be more stable when B-B bonds between octahedra are longer than those in cP7 (Fig. \ref{cP7_oS56_BB}). 
\begin{center}
\begin{table*}[t]
\begin{tabular}{c|ccccccccccc}
\hline\hline
   number     & eigenvector                 & starting  &   final   & Pearson  &  H-H$_{cP7}$  &  V-V$_{cP7}$  &   similarity  &  dynamical     \\
              & combination                 &    SG\#   &    SG\#   &  symbol  &   [meV/atom]  &[atom/\AA$^3$] &     to oP28   &  stability     \\ \hline
     1        &  $e^R_1+e^R_2+e^R_3+e^M_1$  &     14    &     62    &   oP28   &    -10.12     &    -0.107     &    1.0000     &     yes        \\
     2        &  $      e^R_2+e^R_3+e^M_1$  &     62    &     62    &   oP28   &    -10.12     &    -0.107     &    0.9986     &     yes        \\
     3        &  $e^R_1+e^R_2+e^R_3      $  &    167    &    167    &   hR42   &     -8.85     &    -0.093     &    0.9081     &     yes        \\
     4        &  $e^R_1+            e^M_1$  &    127    &    140    &   tP28   &     -9.96     &    -0.108     &    0.6462     &     yes        \\
     5        &  $e^R_1                  $  &    140    &    140    &   tI28   &     -9.96     &    -0.109     &    0.6497     &     yes        \\
     6        &  $                  e^M_1$  &    127    &    127    &   tP14   &     -8.49     &    -0.101     &    0.6262     &     no         \\
     7        &  $e^R_1+e^R_2            $  &     74    &     74    &   oI28   &     -9.07     &    -0.094     &    0.7617     &     no         \\
     8        &  $e^R_1+e^R_2+      e^M_1$  &     12    &     12    &   mS56   &     -9.99     &    -0.107     &    0.7538     &     no         \\
     9        &  $      e^R_2+      e^M_1$  &     63    &     63    &   oS56*   &     -9.99     &    -0.108     &    0.7670     &     no         \\
\hline\hline
\end{tabular}
\caption{Structure and stability of nine possible distorted structures for a 2$\times$2$\times$2 cP7-CaB$_6$ supercell using combinations of four eigenvectors and relaxed at 50 GPa. The table gives the starting and final space group symmetry, Pearson symbol, enthalpy gain and compression with respect to cP7, similarity to oP28 defined through radial distribution functions (Supplemental Information in Ref. \cite{Kolmogorov2012}), and the structures' dynamical stability.  
\newline * Structure 9 is different from the proposed oS56 ground state.}
\label{table_cab6_eigen}
\end{table*}
\end{center}

\vspace{-0.8cm}
At 32 GPa all octahedra open up, rearrange and fuse with three others to form a 24-atom boron unit in a tetragonal tI56 structure.  The evolutionary search revealed a metastable state where one boron atom had migrated to the center of the B$_{24}$ unit.  As there is enough space in the middle for an extra boron atom we constructed unit cells with additional boron atoms.  Filling all of the B$_{24}$ units led to a tI58-CaB$_{6.25}$ phase metastable by 6 meV/atom with respect to $\gamma$-B and tI56-CaB$_6$ at 50 GPa. However, filling just half of the B$_{24}$ units resulted in a stable tP57-CaB$_{6.125}$ phase and made the empty tI56-CaB$_6$ metastable by 5 meV/atom with respect to tP57-CaB$_{6.125}$ and oP20-CaB$_4$ at this pressure.  We only tested the two most ordered configurations of boron atoms inside the units and it is possible that others could result in lower-energy ordered or disordered CaB$_{6+x}$ structures based on the mixture of B$_{24}$ and B$_{25}$ building blocks.  We find that it would be difficult to resolve such derivatives of the tI56-CaB$_6$ phase with powder X-ray measurements due to closeness of the simulated powder XRD patterns.  Out of all alkaline-earth metal hexaborides, the calcium structure is stable as tP57 since it provides the smallest cuboid volume in the B$_{24}$ unit for a boron atom to sit. \cite{supmat}
\begin{figure}[t]
\begin{center}
\includegraphics[width=85mm]{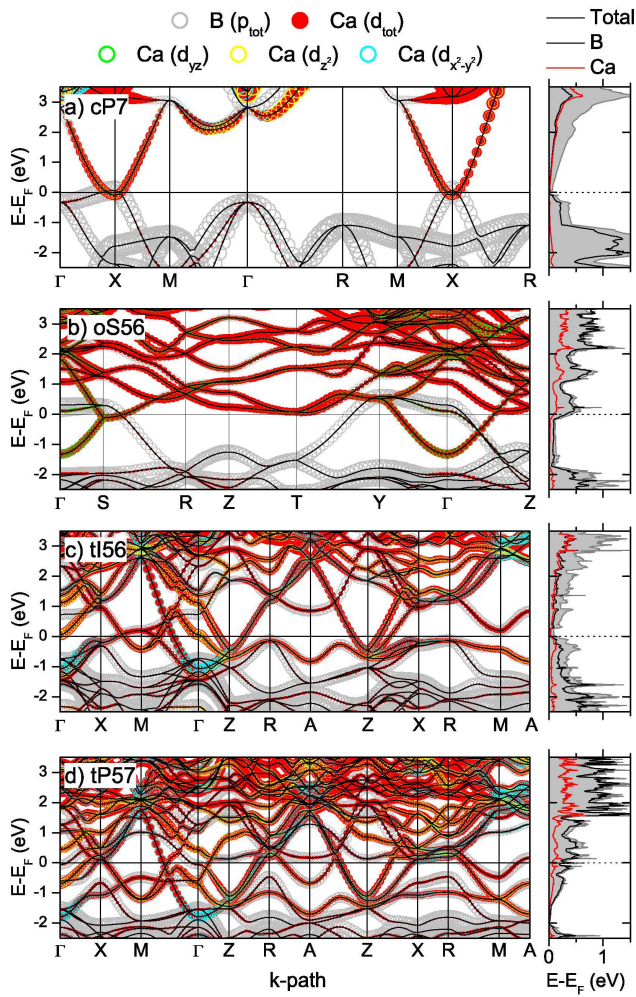}
\caption{(Color online) Calculated band structure and density of states plots of known (cP7-CaB$_6$) and predicted (oS56-CaB$_6$, tI56-CaB$_6$ and tP57-CaB$_{6.125}$) structures relaxed at 30 GPa.  The size of the colored circles indicates the orbital contributions to the bands.  The decomposition was done within the default PAW radius.  The \textit{k}-paths were generated using Refs. \cite{Aroyo2006a,Ayoro2006_2,Setyawan2010}.  The conventional unit cell was used for tI56 to illustrate the differences in the tI56 and tP57 band structures.}
\label{bs_cab6}
\end{center}
\end{figure}

The electronic properties of known cP7 and proposed structures of CaB$_6$ (oS56, tI56) and tP57-CaB$_{6.125}$ were calculated within the GGA (Fig. \ref{bs_cab6}).  These tests were intended only for preliminary analysis of the electronic states because standard DFT incorrectly predicts cP7-CaB$_6$ as being metallic \cite{Lee2005}.  Since there is no significant change in the electronic density of states or band structure for the cP7 structure under pressure, the driving force for the phase transition at 13 GPa is thought to be structural (Sec. \ref{sec:cab6_structure}). 

The decomposed band structure plots in Fig. \ref{bs_cab6} show orbital contributions to specific bands close to the Fermi level. The valence bands close to and below the Fermi level are dominated by boron $p$ states in all three structures.  The conduction band for the cP7 structure across $\Gamma$-$X$-$M$ is composed of a hybrid between B($p$) and Ca($d_{z^2}$) and Ca($d_{x^2-y^2}$).  In the oS56 structure, it is the nearly free-electron Ca($d_{yz}$) states that hybridize with the B-$p$ states near the Fermi level and produce a nearly constant DOS in the pseudogap region from -2 to 0 eV. The Fermi level lies on the slope of a DOS peak further suggesting that this is a suboptimal configuration and a more stable related structure might be found by considering larger unit cells. In the tI56 structure, just as in cP7, the Ca ($d$) states close to the Fermi level are from Ca ($d_{z^2}$) and Ca ($d_{x^2-y^2}$).  The bottom of the Ca $d$ bands in both oS56 and tI56 structures now lies well below the Fermi level (by $\sim$1 eV) while the top of the predominantly B $p$ bands lies above the Fermi level (by $\sim$0.5 eV). These observations suggest that the predicted polymorphs should be metallic with multiple Fermi surfaces of mixed Ca-$d$/B-$p$ or predominantly B-$p$ character but more accurate methods, such as \textit{GW} or time-dependent DFT, should be used to determine the exact position of these electronic states and whether the band gap in cP7 closes at high pressures.

The insertion of an extra boron atom into half of tI56-CaB$_6$ boron units results in mostly boron states at \textit{E$_F$}.  These states are mainly B($p_y$) which lie in the direction of the filled boron unit layers.  The Ca($d_{x^2-y^2}$) band crossing $\Gamma$ around -1 eV in tI56 is moved down by 0.7 eV in tP57 due to the filling of hybrid B($p$) and Ca($d_{z^2}$) band.     

\subsection{Superconducting properties of CaB$_6$}

We predict that the metastable tI58-CaB$_{6.25}$ would be superconducting at 50 GPa with a \textit{T$_c$} of 1.9 K ($\lambda$ = 0.41, ${\omega_{ln}}$ = 653.9 cm$^{-1}$).  tI56-CaB$_6$ is predicted to be non-superconducting.  Due to the large structure size, we could not complete the calculation of the Eliashberg function for the more stable tP57-CaB$_{6.125}$.  However, judging by the DOS at \textit{E$_F$} one could expect the structure to be superconducting with \textit{T$_c$} of $~$1 K \cite{supmat}.  

\section{C\lowercase{a}B$_4$} \label{sec:CaB4}
Twenty metal tetraborides are known to have a common structure, the tetragonal ThB$_4$-type tP20 structure, first determined by Zalkin and Templeton \cite{Zalkin1953} in 1952.  Tetraborides are of interest for their electronic (e.g. LaB$_4$) \cite{Lafferty1950} and magnetic properties (e.g. \textit{M}B$_4$ where \textit{M} = Nd, Sm, Er and Yb). \cite{Yin2008}  Calcium tetraboride is the only alkaline-earth metal thought to form this structure and although it was first synthesized in 1961, \cite{Johnson1961} the compound has regained some interest recently. \cite{Schmitt2006, Yahia2008, Yin2008, Liu2010} {\color{black} Previous DFT-based calculations have found tP20-CaB$_4$ to be stable at ambient pressure. \cite{Chepulskii2009}  Liu \textit{et al.} \cite{Liu2010} recently reported the experimental synthesis of CaB$_4$ under 2 GPa of pressure, however from computational analysis they concluded that the formation of CaB$_4$ is not favorable under ambient pressure in comparison to CaB$_6$.}

\begin{figure}[!ht]
\begin{center}
\includegraphics[width=60mm]{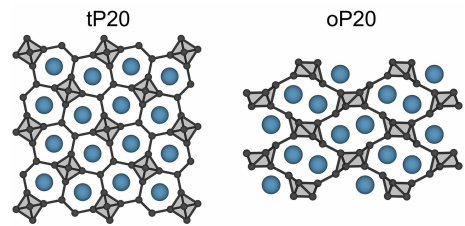}
\caption{(Color online) A top view of one layer of the crystal structure of the tP20-CaB$_4$ ground state at 0 GPa and a side view of the oP20 MgB$_4$-type structure of CaB$_4$, a proposed ground state at 30 GPa.}
\vspace{-0.3cm}
\label{CaB4_structures}
\end{center}
\end{figure}

The ThB$_4$ type structure (Fig. \ref{CaB4_structures}) consists of a boron layer with B$_6$ octahedra, as seen in \textit{M}B$_6$, \cite{Mutterities1967}  joined by B$_2$ dimers as in each unit cell of \textit{M}B$_2$\cite{Mutterities1967}.  Metal atoms sit above and below the boron layer and B-B bonds join the layers of octahedra forming a three-dimensional covalent boron network.  It has been suggested that insufficient electrons are provided by the calcium ion to the boron network. \cite{Schmitt2006, Yahia2008} For CaB$_4$ to be stable, the dimer must be a double bond \cite{Burdett1991} taking two electrons from a Ca$^{2+}$ ion.  The rarity of a boron-boron double bond has led to the conclusion that three electrons are required from the metal atom for the stability of a single bonded dimer. \cite{Schmitt2006, Yahia2008} However, it has been shown that semi-filled bonding bands are sufficient for the stability of lanthanide and actinide tetraborides. \cite{Yin2008}  It is therefore clear that some questions about the stability and electronic structure of CaB$_4$ still remain unanswered.  In addition, to our knowledge there is no evidence of a pressure-induced phase transition from ThB$_4$-type calcium tetraboride.  

Transition metal tetraboride structures (e.g. CrB$_4$, MoB$_4$ and WB$_4$) are not relevant for this study as their alkaline-earth analogs are unstable.  It is however interesting to note that even for other tetraborides there have been many open questions at ambient pressure.  For example, we predicted that the ground state crystal structure of the known CrB$_4$ compound has lower symmetry \cite{bialon2011} than thought originally \cite{Andersson1968} and our proposed structure type has been subsequently confirmed with electron diffraction experiments. \cite{Niu2012}

Here we explain why ThB$_4$-type CaB$_4$ is stable at ambient pressure and give a general rule for the stability of all ThB$_4$-type tetraborides with pressure.  Stability is shown to be a balance between the size restrictions posed by the metal ion within the boron lattice and the distribution of charge between B$_6$ and B$_2$ units. Using an evolutionary search we found that at 19 GPa the oP20 MgB$_4$-type structure becomes energetically favorable relative to the low-pressure tP20 ThB$_4$ type (Fig. \ref{CaB4_structures}).  oP20-CaB$_4$ is composed of a chain of pentagonal pyramids with metal atoms sitting in channels running in one direction.  At 30 GPa, the tP20 structure is 39 meV/atom higher in enthalpy than the proposed oP20 ground state (Fig. \ref{fig1}).  Two higher pressure (50 GPa) semiconducting metastable states (mP20 and hR10) were also revealed in the search as detailed in the Supplemental Material. \cite{supmat}   

\subsection{Structural stability of CaB$_4$}
\begin{figure}[t]
\begin{center}
\includegraphics[width=70mm]{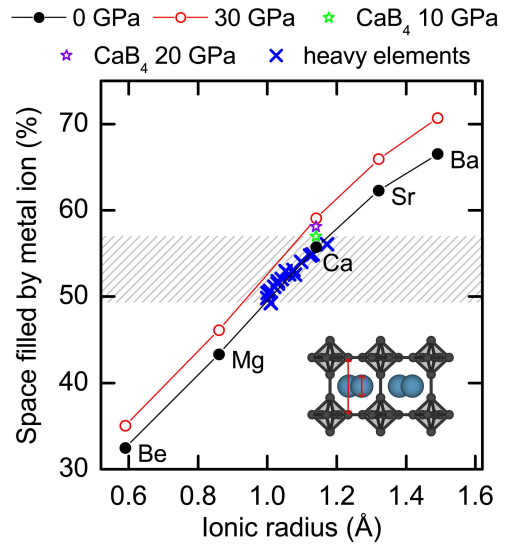}
\caption{(Color online) The space filled by the metal ion in the c axis plotted against the ionic radius, for the tP20 structure of \textit{M}B$_4$.  The space filled is the ratio between the smaller and larger red arrows on the crystal structure diagram.  The gray shaded region (49-57\%) shows the space filling corresponding to a stable tP20 structure.  Heavy elements are these lanthanides and actinides: Th, Lu, Ce, Nd, Y, Gd, Sm, Tb, Er, Pr, Dy, La, Pu, Ho, U and Tm.  Alkaline-earth metals are assumed to be divalent.  Lanthanides are assumed to be trivalent as known in the literature. \cite{Yin2008}}
\label{tP20}
\end{center}
\end{figure}

Metal ions sit directly on top of each other in tP20-CaB$_4$ and since these ions are closer together than those in the plane, one can expect this interplanar separation to have an effect on stability.  Figure \ref{tP20} shows how the percentage space filled up by the divalent metal ion perpendicular to the boron plane varies with ionic radius and pressure for the alkaline-earth metal borides.  Going down the group, the interplanar separation (lattice vector \textit{c}) does not increase greatly to accommodate the larger ionic radius so the space filled up by the ion increases almost linearly.  As expected, applying pressure reduces the space available to the ion but only slightly as the rigidity of the boron lattice in three dimensions is maintained by the B-B bond between the layers.  All known metal tetraborides with the tP20 structure fit neatly over a small range where 49-57\% of the interplanar distance can be filled with the metal ion.  At both 0 and 30 GPa, MgB$_4$ falls below this range which is the reason for the tP20 structure being 105 and 293 meV/atom higher in enthalpy than the ground state at these two pressures.  The experimentally observed tetraborides of the heavy elements, lanthanides and actinides, are plotted here and all fall within this range.  CaB$_4$ is stable in the tP20 structure up to 19 GPa before becoming thermodynamically unstable relative to the oP20 structure.  tP20-CaB$_4$ at 20 GPa lies outside the proposed allowed filling range.  At high space filling, direct metal-metal repulsion makes this tP20 configuration thermodynamically unfavorable. 

At ambient pressure, tetraborides with metallic ions of radii 0.99-1.18 {\AA} are predicted to form the tP20 structure.  This prediction can be used to explain the formation of tetraborides for all metals.  Muetterties \cite{Mutterities1967} defined an atomic radius range over which tetraborides were stable (1.97 - 1.68 {\AA}, center - 1.82 {\AA}) but we have defined it in terms of an ionic radius.  This is because: (i) charge transfer from the metal to the boron network is known to occur and (ii) the ion space filling is more strongly correlated to the compounds' stability than atomic space filling. \cite{supmat}

Monovalent sodium is the only alkali metal that falls within the range; however, as two electrons are not available to be transferred to the boron network, sodium does not form a stable structure, lying 120 meV/atom above the Na$_3$B$_{20}$-Na tie-line.  RbB$_4$ has been suggested to stabilize as tP20 \cite{Chepulskii2009} but our calculations indicate that it is unstable with a positive formation enthalpy of 253 meV/atom.  Across the row of \textit{3d} transition metals, only one ion (1+, 2+, 3+, 4+ or 5+), Ti$^{2+}$, fits within the range.  Since titanium prefers to be in the 3+ or 4+ state, tP20-TiB$_4$ does not form.  Among \textit{4d} transition metals, yttrium with an ionic radius of 1.04 {\AA} falls within the window and forms a stable tP20 tetraboride.  Divalent palladium, silver and cadmium have ionic radii of 1.00, 1.08 and 1.09 {\AA}, respectively, falling within the stability window.  The tP20 structure is not stable for these borides as the metals are thought to prefer to be in other valence states (for example silver prefers to be monovalent).  No cadmium borides are known and palladium only forms metal rich borides.  Praseodymium and europium tetraborides are the only lanthanide tetraborides that are not stable at ambient pressure. \cite{Yin2008}  Trivalent praseodymium has an ionic radius of 1.13 {\AA}, so theoretically we would expect this to form a tP20 structure; however, we are unlikely to see any experimental studies to prove this due to the low abundance of the element.  Trivalent europium has an ionic radius of 1.09 {\AA}; however, it is thought that the element prefers to be in a 2+ state which would make the ion too large (1.31 {\AA}) to form the tP20 tetraboride structure.  All the known actinide tetraborides fall within the window (Th, U, Np, Pu, Am).  The trivalent ions of Pa, Cm, Bk and Cf are of an appropriate size to fit into the window however the rarity of these elements makes it difficult to form a sound conclusion as to whether or not they will form a tetraboride.  

The phase transition from tP20 to oP20 makes CaB$_4$ more compact: at 30 GPa the volume per ion for oP20 is 8.45 {\AA}, while for tP20 it is 8.97 {\AA}.  Naslain {\itshape et al.} \cite{Naslain1973} studied the structure of magnesium tetraboride and explained that larger ions prefer to be in the tP20 structure while the smaller magnesium ion prefers the oP20 structure.  We extend this idea to include the effect of externally applied pressure and to explain why the oP20 structure destabilizes at high pressures for all alkaline-earths.  We find that there is an ideal volume for the oP20 structure to be stable, 7.78 - 9.1 {\AA}$^3$ per atom. \cite{supmat}  
\subsection{Electronic properties of CaB$_4$}
Examination of the charge on the B$_2$ dimer in various tP20-type metal tetraborides could give insights into their stability.  Unfortunately, there is no universal definition of an atomic charge in the DFT.  Bader charge decomposition \cite{Tang2009} has revealed interesting charge transfer trends in several systems. \cite{Zhang2012, Niu2012}  We have used the highest accuracy settings (dense \textit{k}-meshes and high cutoff energies) to calculate the charge distribution in tP20-CaB$_4$ as a function of pressure.  We observed that at pressures above 20 GPa, the excess Bader charge on the dimer (relative to three electrons per boron) falls below 0.4 electrons. \cite{supmat}  In comparison, the excess charge on the octahedron does not change much with pressure.   Low charge around the dimer could be thought to lead to instability of the B$_2$ unit and thus the whole tP20 framework.  Using this boundary one could also predict destabilization of YB$_4$ and LaB$_4$ above 50 GPa. \cite{supmat}  

\begin{figure}[t]
\begin{center}
\includegraphics[width=8cm]{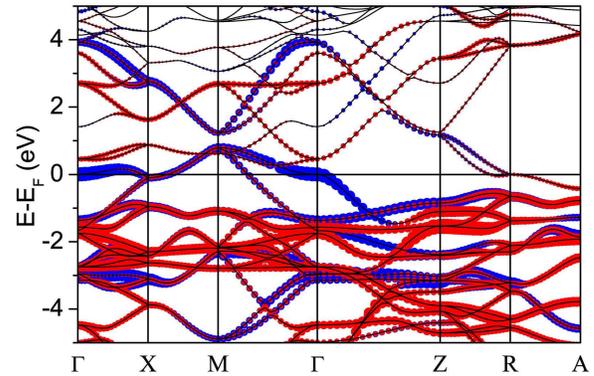}
\caption{(Color online) The electronic band structure of tP20-CaB$_4$ at 0 GPa.  Blue and red circles show the B$_2$ and B$_6$ bands respectively.}
\label{tP20_0GPa_band}
\end{center}
\end{figure}

The electronic band structure shows that tP20-CaB$_4$ is metallic (Fig. \ref{tP20_0GPa_band}).  Unlike the other Ca-B structures discussed in this study, the DOS at the Fermi level is dominated by boron $p$ states.  The semi-filled bands around the Fermi level are mainly from the B$_2$ dimer.  The dimer bands have also been found to be semi-filled in stable lanthanide and actinide tetraborides. \cite{Yin2008} Our GGA calculations indicate that the high pressure oP20 structure is semiconducting.        

\section{C\lowercase{a}B$_2$} \label{sec:CaB2}
The diborides have become the most studied boride composition after the discovery of superconductivity in MgB$_2$, \cite{Buzea2001} though among the alkaline-earths, only magnesium diboride is stable at ambient pressure.  A stable phase of BeB$_2$ has previously been proposed. \cite{Hermann2012, Sands1961}  However, the experiments were not able to fully characterize the structure.  Furthermore, our calculations indicate that the theoretically determined \textit{F$\bar{4}$3m} structure proposed \cite{Hermann2012} is thermodynamically unstable (by 84 meV/atom) relative to at least one structure found in our preliminary evolutionary search of buckled hexagonal sheets of boron with a double layer of metal ions. \cite{supmat}  A CaB$_2$ structure of linear boron chains was recently predicted to stabilize at ambient pressure. \cite{Vajeeston2012}  This structure was previously proposed ($\delta$-CaB$_2$) and found to be highly unstable against phase separation into CaB$_6$ and fcc-Ca by 144 meV/atom due to the large ionic size of calcium. \cite{Kolmogorov2006}

\raggedbottom

\begin{figure}[!ht]
\begin{center}
\includegraphics[width=70mm]{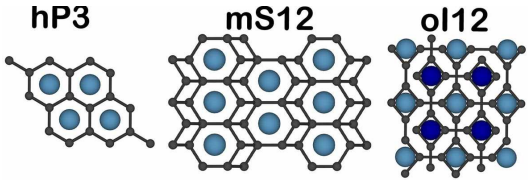}
\caption{(Color online) A top view of the planes that form the crystal structures of three considered CaB$_2$ ground states at 30 GPa: the AlB$_2$-type hP3 structure and proposed mS12 and oI12 structures.  The dark blue in oI12 represents the calcium atom in the next layer which is shifted relative to the top layer.}
\label{CaB2_structures3}
\end{center}
\end{figure}

At high pressure, MgB$_2$ is known to be stable in the hP3-AlB$_2$ (hP3) configuration up to at least 57 GPa. \cite{Goncharov2003}  As can be seen in Fig. \ref{CaB2_structures3}, the hP3 structure consists of hexagonal layers of boron sandwiching closed-packed layers of metal atoms.  It has been suggested that CaB$_2$ stabilizes as hP3 above 8 GPa. \cite{Vajeeston2012}  Our calculations indicate that at 30 GPa, hP3-CaB$_2$ lies 56 meV/atom above the CaB$_4$-Ca tie-line (see Fig. \ref{fig1}).  We propose a different superconducting oI12 structure above 8 GPa, consisting of ABA stacking of B-Ca layers with four and eight membered rings of boron.  The 4-8 arrangement of the covalent boron network makes room for the large Ca ion to sit over the eight-membered ring.  At 32 GPa a monoclinic structure (mS12) of buckled layers of boron hexagons becomes thermodynamically more stable.  

\subsection{Structural properties and stability of potential ground states of CaB$_2$}
hP3-CaB$_2$ is unstable at ambient pressure as metal ions are too close to each other in the plane.  We have found more energetically favorable configurations of CaB$_2$ at 0 GPa.  The lowest energy structure is the proposed high-pressure phase of oI12-CaB$_2$.  At 30 GPa, the formation enthalpy of CaB$_2$ structures is generally found to be dependent on the shortest metal-boron distance (M-B) within the structure. \cite{supmat}  The M-B in oI12 is 2.61 {\AA}, which was the largest of all the structures considered.  The eight-membered boron ring in oI12, restricts the metal ion to sit above the center of such a large ring, resulting in a large M-B.  

\subsection{Electronic properties of CaB$_2$}
\begin{figure}[!h]
\begin{center}
\includegraphics[width=8cm]{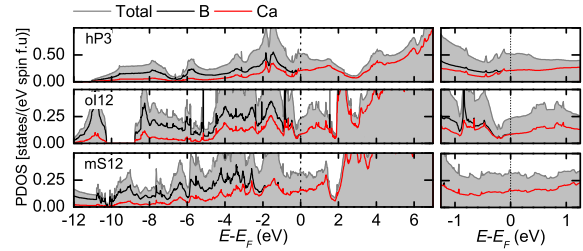}
\caption{(Color online) Projected density of states of three structures of CaB$_2$ relaxed to 30 GPa.  PDOS is stated in units of states per eV per spin per formula unit (f.u.) against the energy relative to the Fermi energy (\textit{E$_F$}).  On the right there is a close up of the states around the Fermi level.  The scale for hP3 is double that for oI12 and mS12.}
\label{CaB2_PDOS}
\end{center}
\end{figure}

The electronic DOS for the unstable hP3 structure and two proposed ground state structures oI12 and mS12 are shown in Fig. \ref{CaB2_PDOS}.  All three structures are metallic with the DOS at the Fermi level for the hP3 structure being around double that for the ground states, illustrating the basic problem of combining desired superconductivity and stability features in one structure.  The hP3 and oI12 structures have equal contributions from both boron and calcium atoms at the Fermi level, while the mS12 structure has a slightly higher contribution from the calcium atom.  Boron states are mainly from $p_y$ and $p_z$ orbitals which lie perpendicular and parallel to the boron plane.  Calcium states are mainly from $d_{xy}$ and $d_{xz}$ orbitals.  

\subsection{Superconducting properties of CaB$_2$}
oI12-CaB$_2$ at 30 GPa and mS12-CaB$_2$ at 50 GPa are expected to be poor superconductors with a \textit{T$_c$} of 0.8 and 1.0 K respectively.  Compared to CaB, which will be discussed next, oI12-CaB$_2$ has a higher $\left\langle\omega_{ln}\right\rangle$ but its DOS at \textit{E$_F$} is under half of that in CaB resulting in a lower $\lambda$.  From a structure search of Sr$_4$B$_8$ we find oI12-SrB$_2$ to be stable at 30 GPa, and predict it to be superconducting with a \textit{T$_c$} of 0.7 K.  Despite Sr having a larger mass than Ca, therefore lowering its $\left\langle\omega_{ln}\right\rangle$, the structure has a similar predicted \textit{T$_c$}.

\section{C\lowercase{a}B}  \label{sec:CaB}
Monoboride structures are unknown among the alkaline-earth metals at both ambient and high pressure as the number of electrons in the compound is not optimal.  Using the results of an evolutionary search we find that CaB stabilizes at 20 GPa as zigzagged boron chains (oI8) and a metastable state of snake like boron chains becomes thermodynamically more stable above 35 GPa.  The two structures are also stable for strontium boride.  
    
\subsection{Structural properties of CaB}
\begin{figure}[!h]
\begin{center}
\includegraphics[width=7cm]{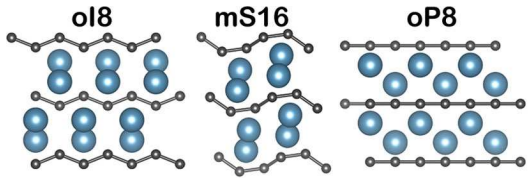}
\caption{(Color online) The crystal structures of the three lowest enthalpy phases of CaB found using the evolutionary search.  The oI8 ground state consisting of zig-zag boron layers, mS16 metastable state consisting of bent boron chains and the linear chained oP8 structure that is the highest in enthalpy out of the three structures.}
\label{CaB_structures}
\end{center}
\end{figure}

With only one boron atom per metal atom there are limited configurations that a covalent boron network can form. \cite{Etourneau1985}  All involve individual chains since a slightly higher concentration of boron is required for double chains (as in Cr$_3$B$_4$) and a slightly lower concentration is required for branched chains (as in Ru$_{11}$B$_8$). \cite{Etourneau1985}  The evolutionary search revealed three low-energy types of boron chains that we have compared: oP8, linear; mS16, snake-like and oI8 - zigzag (Fig. \ref{CaB_structures}).  oI8-CaB is an ABA layered structure with calcium atoms sitting in the bends of the chains.  Linear chained oP8 is 53 meV/atom higher in enthalpy than oI8.   Boron chains fit themselves around group II metal ions, therefore, the larger the ion, the larger the bond angle in the zigzagged structure (oI8).  Comparing to the linear chains (oP8) we find that the smallest ions (Be and Mg) relax away from linear chains and back to zigzag chains (oI8).  Calcium boride relaxes as linear chains only when 20 GPa of pressure is applied.    

\subsection{Electronic properties of CaB}
\begin{figure}[!ht]
\begin{center}
\includegraphics[width=80mm]{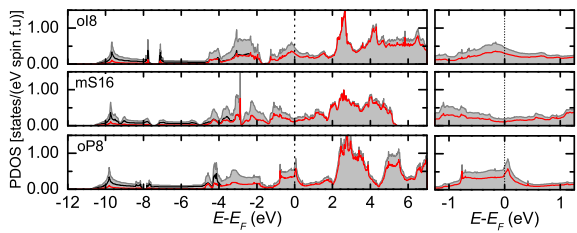}
\caption{(Color online) The projected density of states of the three lowest enthalpy structures (oI8, mS16 and oP8) of CaB found using the evolutionary search.  See Fig. \ref{CaB2_PDOS} for the graph labels.}  
\label{PDOS-CaB}
\end{center}
\end{figure}

The electronic density of states (Fig. \ref{PDOS-CaB}) shows the metallic nature of all the considered CaB structures.  The DOS at Fermi level of oI8 and oP8 are 0.45 and 0.65 states/(eV spin f.u.), respectively, coming mainly from the calcium atom, in particular Ca($d_{x^{2}-y^{2}}$) states which lie along the line of the boron chains and between the layers.  Boron $p_y$ and $p_z$ orbitals which lie in the zig-zag plane contribute most to boron states at the Fermi level.  The mS16 structure has a much lower DOS at \textit{E$_F$} [0.20 states/(eV spin f.u.)], possibly resulting from a higher contribution (50\%) to the total DOS from the boron.

\subsection{Superconducting properties of CaB}
\begin{figure}[!ht]
\begin{center}
\includegraphics[width=8.0cm]{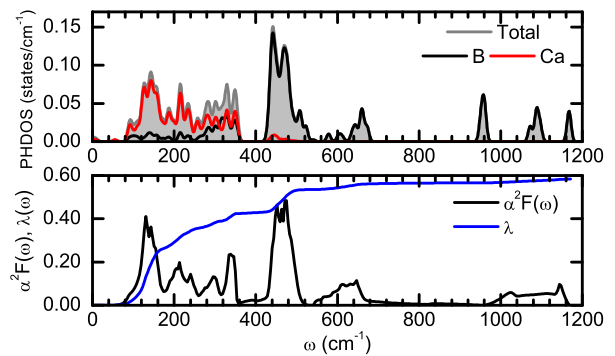}
\caption{(Color online) Top: total and projected phonon density of states (PHDOS), bottom: Eliashberg function \textit{$\alpha^{2}$F($\omega$)} and $\lambda$ of oI8-CaB at 30 GPa.} 
\label{oI8-CaB-ph-all}
\end{center}
\end{figure}

Figure \ref{oI8-CaB-ph-all} shows the Eliashberg spectral function and the electron-phonon coupling strength $\lambda$ for oI8-CaB.  We predict that oI8-CaB will be superconducting at 30 GPa with a \textit{T$_c$} of 5.5 K, total $\lambda$ of 0.58 and $\left\langle\omega_{ln}\right\rangle$ of 330.2 cm$^{-1}$.  68$\%$ of the total $\lambda$ results from modes below 360 cm$^{-1}$, which are mainly displacements of calcium atoms.  The decomposition of the phonon density of states into contributions from the atoms (Fig. \ref{oI8-CaB-ph-all}) shows that the in-plane movement of boron atoms (around 500 cm$^{-1}$) has a negligible contribution to the electron-phonon coupling in this compound.  {\color{black} Similarly, a calcium phonon mode is thought to play a large role in the linear increase of the \textit{T$_c$} of CaC$_6$ with pressure (11.5 K at 0 GPa up to a maximum of 15.1 K at 7.5 GPa). \cite{Gauzzi2007}}  It would also be interesting to compare the superconducting properties of the CaB and the related isoelectronic oS8-LiC compounds.  The latter has been recently predicted to form under pressure. \cite{Chen2010}

\vspace {-0.0cm}

\section{Conclusions}
We have used an evolutionary structure search to identify high-pressure stable phases of Ca$_x$B$_{1-x}$.  At 30 GPa three new structures of CaB$_6$, CaB$_2$ and CaB stabilize, while CaB$_4$ stabilizes as the known MgB$_4$ structure type.  For all compositions we have found that the stability of a particular structure is highly dependent on the size of the metal ion, shaping the boron framework around it.  

A variety of stable superconducting calcium boride compounds are predicted but have relatively low critical temperatures, exemplifying the inverse correlation between stability and superconducting \textit{T$_c$}.  Stable superconductors are only found for medium concentrations of calcium where $x$ = 0.33-0.5.  Their superconducting nature is thought to be due to the high contribution of calcium states at the Fermi level.  The main results for each Ca-B composition are summarized below.     

CaB$_6$: The known semiconducting cubic structure (cP7) at ambient pressure becomes dynamically unstable around 25 GPa.  Three derived dynamically stable structures have been identified via a systematic analysis of multiple imaginary-frequency phonon modes.  The stability of a metallic orthorhombic structure (oS56), proposed to be the CaB$_6$ ground state in the 13-32 GPa pressure range, \cite{Kolmogorov2012} is shown to correlate with the length of a boron bond in all alkaline-earth hexaborides (Fig. \ref{cP7_oS56_BB}).  The parent tI56-CaB$_6$ structure, predicted to be stable above 32 GPa and supported by XRD data in our previous study, \cite{Kolmogorov2012} is shown here to stabilize further with the addition of a small amount of boron.  The more stable tP57-CaB$_{6.125}$ structure derived from tI56-CaB$_6$ may be a low-\textit{T$_c$} superconductor.

CaB$_4$: The ThB$_4$-type structure (tP20) is demonstrated to be thermodynamically stable at ambient pressure.  A general rule has been found, for all pressures, relating the stability of tP20-\textit{M}B$_4$ with the size of the metal ion (Figure \ref{tP20}).  According to the rule, at ambient pressure, metals with ionic radii of 0.99-1.18 {\AA} should form tP20-\textit{M}B$_4$.  tP20-CaB$_4$ destabilizes at 19 GPa relative to the MgB$_4$-type semiconducting structure.

CaB$_2$: No stable polymorphs have been found at ambient pressure but a four- and eight-membered boron ring layered metallic structure (oI12) is expected to form under 8 GPa of pressure.  oI12-CaB$_2$ and its SrB$_2$ analog are superconducting at $\sim$1 K.  Above 32 GPa, a monoclinic structure (mS12) composed of bucked sheets of hexagons, which is poorly superconducting, becomes thermodynamically more stable. 

CaB: No alkaline-earth monoborides are known at ambient pressure.  Above 20 GPa, a superconducting zigzag chained structure of oI8-CaB stabilizes with a predicted \textit{T$_c$} of 5.5 K.  The boron chains are zigzagged as they fit themselves around the metal ion.    

\vspace{0.5cm} 
\noindent We acknowledge the support of the EPSRC CAF EP/G004072/1.
\newpage
\bibliography{refs}

\begin{thebibliography}{83}
\expandafter\ifx\csname natexlab\endcsname\relax\def\natexlab#1{#1}\fi
\expandafter\ifx\csname bibnamefont\endcsname\relax
  \def\bibnamefont#1{#1}\fi
\expandafter\ifx\csname bibfnamefont\endcsname\relax
  \def\bibfnamefont#1{#1}\fi
\expandafter\ifx\csname citenamefont\endcsname\relax
  \def\citenamefont#1{#1}\fi
\expandafter\ifx\csname url\endcsname\relax
  \def\url#1{\texttt{#1}}\fi
\expandafter\ifx\csname urlprefix\endcsname\relax\def\urlprefix{URL }\fi
\providecommand{\bibinfo}[2]{#2}
\providecommand{\eprint}[2][]{\url{#2}}

\bibitem[{\citenamefont{Curtarolo et~al.}(2013)\citenamefont{Curtarolo, Hart,
  Nardelli, Mingo, Sanvito, and Levy}}]{Curtarolo2013}
\bibinfo{author}{\bibfnamefont{S.}~\bibnamefont{Curtarolo}},
  \bibinfo{author}{\bibfnamefont{G.~L.~W.} \bibnamefont{Hart}},
  \bibinfo{author}{\bibfnamefont{M.~B.} \bibnamefont{Nardelli}},
  \bibinfo{author}{\bibfnamefont{N.}~\bibnamefont{Mingo}},
  \bibinfo{author}{\bibfnamefont{S.}~\bibnamefont{Sanvito}}, \bibnamefont{and}
  \bibinfo{author}{\bibfnamefont{O.}~\bibnamefont{Levy}}, \bibinfo{journal}{Nat
  Mater} \textbf{\bibinfo{volume}{12}}, \bibinfo{pages}{191}
  (\bibinfo{year}{2013}).

\bibitem[{\citenamefont{Hautier et~al.}(2012)\citenamefont{Hautier, Jain, and
  Ong}}]{Hautier2012}
\bibinfo{author}{\bibfnamefont{G.}~\bibnamefont{Hautier}},
  \bibinfo{author}{\bibfnamefont{A.}~\bibnamefont{Jain}}, \bibnamefont{and}
  \bibinfo{author}{\bibfnamefont{S.~P.} \bibnamefont{Ong}},
  \bibinfo{journal}{Journal of Materials Science}
  \textbf{\bibinfo{volume}{47}}, \bibinfo{pages}{7317} (\bibinfo{year}{2012}).

\bibitem[{\citenamefont{Eliashberg}(1960)}]{Eliashberg1960}
\bibinfo{author}{\bibfnamefont{G.}~\bibnamefont{Eliashberg}},
  \bibinfo{journal}{Soviet Physics - JETP} \textbf{\bibinfo{volume}{11}},
  \bibinfo{pages}{696} (\bibinfo{year}{1960}).

\bibitem[{\citenamefont{Allen and Dynes}(1975)}]{Allen1975}
\bibinfo{author}{\bibfnamefont{P.~B.} \bibnamefont{Allen}} \bibnamefont{and}
  \bibinfo{author}{\bibfnamefont{R.~C.} \bibnamefont{Dynes}},
  \bibinfo{journal}{Physical Review B} \textbf{\bibinfo{volume}{12}},
  \bibinfo{pages}{905} (\bibinfo{year}{1975}).

\bibitem[{\citenamefont{Margine and Giustino}(2013)}]{Margine2013}
\bibinfo{author}{\bibfnamefont{E.~R.} \bibnamefont{Margine}} \bibnamefont{and}
  \bibinfo{author}{\bibfnamefont{F.}~\bibnamefont{Giustino}},
  \bibinfo{journal}{Physical Review B} \textbf{\bibinfo{volume}{87}},
  \bibinfo{pages}{024505} (\bibinfo{year}{2013}).

\bibitem[{\citenamefont{Cohen}(2011)}]{CohenBCS}
\bibinfo{author}{\bibfnamefont{M.}~\bibnamefont{Cohen}}, in
  \emph{\bibinfo{booktitle}{BCS: 50 Years}}, edited by
  \bibinfo{editor}{\bibfnamefont{L.}~\bibnamefont{Cooper}} \bibnamefont{and}
  \bibinfo{editor}{\bibfnamefont{D.}~\bibnamefont{Feldman}}
  (\bibinfo{publisher}{World Scientific}, \bibinfo{address}{Singapore},
  \bibinfo{year}{2011}).

\bibitem[{\citenamefont{Chang and Cohen}(1984)}]{Chang1984}
\bibinfo{author}{\bibfnamefont{K.~J.} \bibnamefont{Chang}} \bibnamefont{and}
  \bibinfo{author}{\bibfnamefont{M.~L.} \bibnamefont{Cohen}},
  \bibinfo{journal}{Phys. Rev. B} \textbf{\bibinfo{volume}{30}},
  \bibinfo{pages}{5376} (\bibinfo{year}{1984}).

\bibitem[{\citenamefont{Neaton and Ashcroft}(1999)}]{Neaton1999}
\bibinfo{author}{\bibfnamefont{J.~B.} \bibnamefont{Neaton}} \bibnamefont{and}
  \bibinfo{author}{\bibfnamefont{N.~W.} \bibnamefont{Ashcroft}},
  \bibinfo{journal}{Nature} \textbf{\bibinfo{volume}{400}},
  \bibinfo{pages}{141} (\bibinfo{year}{1999}).

\bibitem[{\citenamefont{Shimizu et~al.}(2002)\citenamefont{Shimizu, Ishikawa,
  Takao, Yagi, and Amaya}}]{Shimizu2002}
\bibinfo{author}{\bibfnamefont{K.}~\bibnamefont{Shimizu}},
  \bibinfo{author}{\bibfnamefont{H.}~\bibnamefont{Ishikawa}},
  \bibinfo{author}{\bibfnamefont{D.}~\bibnamefont{Takao}},
  \bibinfo{author}{\bibfnamefont{T.}~\bibnamefont{Yagi}}, \bibnamefont{and}
  \bibinfo{author}{\bibfnamefont{K.}~\bibnamefont{Amaya}},
  \bibinfo{journal}{Nature} \textbf{\bibinfo{volume}{419}},
  \bibinfo{pages}{597} (\bibinfo{year}{2002}).

\bibitem[{\citenamefont{Pickett}(2008)}]{Pickett2008}
\bibinfo{author}{\bibfnamefont{W.~E.} \bibnamefont{Pickett}},
  \bibinfo{journal}{Physica C: Superconductivity}
  \textbf{\bibinfo{volume}{468}}, \bibinfo{pages}{126} (\bibinfo{year}{2008}).

\bibitem[{\citenamefont{Buzea and Yamashita}(2005)}]{Buzea2005}
\bibinfo{author}{\bibfnamefont{C.}~\bibnamefont{Buzea}} \bibnamefont{and}
  \bibinfo{author}{\bibfnamefont{T.}~\bibnamefont{Yamashita}},
  \bibinfo{journal}{Superconductor Science and Technology}
  \textbf{\bibinfo{volume}{18}} (\bibinfo{year}{2005}).

\bibitem[{\citenamefont{Lortz et~al.}(2006)\citenamefont{Lortz, Wang, Tutsch,
  Abe, Meingast, Popovich, Knafo, Shitsevalova, Paderno, and
  Junod}}]{Lortz2006}
\bibinfo{author}{\bibfnamefont{R.}~\bibnamefont{Lortz}},
  \bibinfo{author}{\bibfnamefont{Y.}~\bibnamefont{Wang}},
  \bibinfo{author}{\bibfnamefont{U.}~\bibnamefont{Tutsch}},
  \bibinfo{author}{\bibfnamefont{S.}~\bibnamefont{Abe}},
  \bibinfo{author}{\bibfnamefont{C.}~\bibnamefont{Meingast}},
  \bibinfo{author}{\bibfnamefont{P.}~\bibnamefont{Popovich}},
  \bibinfo{author}{\bibfnamefont{W.}~\bibnamefont{Knafo}},
  \bibinfo{author}{\bibfnamefont{N.}~\bibnamefont{Shitsevalova}},
  \bibinfo{author}{\bibfnamefont{Y.~B.} \bibnamefont{Paderno}},
  \bibnamefont{and} \bibinfo{author}{\bibfnamefont{A.}~\bibnamefont{Junod}},
  \bibinfo{journal}{Phys. Rev. B} \textbf{\bibinfo{volume}{73}},
  \bibinfo{pages}{024512} (\bibinfo{year}{2006}).

\bibitem[{\citenamefont{Flores-Livas et~al.}(2012)\citenamefont{Flores-Livas,
  Amsler, Lenosky, Lehtovaara, Botti, Marques, and
  Goedecker}}]{Flores-Livas2012}
\bibinfo{author}{\bibfnamefont{J.~A.} \bibnamefont{Flores-Livas}},
  \bibinfo{author}{\bibfnamefont{M.}~\bibnamefont{Amsler}},
  \bibinfo{author}{\bibfnamefont{T.~J.} \bibnamefont{Lenosky}},
  \bibinfo{author}{\bibfnamefont{L.}~\bibnamefont{Lehtovaara}},
  \bibinfo{author}{\bibfnamefont{S.}~\bibnamefont{Botti}},
  \bibinfo{author}{\bibfnamefont{M.~A.~L.} \bibnamefont{Marques}},
  \bibnamefont{and}
  \bibinfo{author}{\bibfnamefont{S.}~\bibnamefont{Goedecker}},
  \bibinfo{journal}{Phys. Rev. Lett.} \textbf{\bibinfo{volume}{108}},
  \bibinfo{pages}{117004} (\bibinfo{year}{2012}).

\bibitem[{\citenamefont{Kolmogorov et~al.}(2008)\citenamefont{Kolmogorov,
  Calandra, and Curtarolo}}]{Kolmogorov2008}
\bibinfo{author}{\bibfnamefont{A.~N.} \bibnamefont{Kolmogorov}},
  \bibinfo{author}{\bibfnamefont{M.}~\bibnamefont{Calandra}}, \bibnamefont{and}
  \bibinfo{author}{\bibfnamefont{S.}~\bibnamefont{Curtarolo}},
  \bibinfo{journal}{Physical Review B} \textbf{\bibinfo{volume}{78}},
  \bibinfo{pages}{094520} (\bibinfo{year}{2008}).

\bibitem[{\citenamefont{An and Pickett}(2001)}]{An2001}
\bibinfo{author}{\bibfnamefont{J.~M.} \bibnamefont{An}} \bibnamefont{and}
  \bibinfo{author}{\bibfnamefont{W.~E.} \bibnamefont{Pickett}},
  \bibinfo{journal}{Phys. Rev. Lett.} \textbf{\bibinfo{volume}{86}},
  \bibinfo{pages}{4366} (\bibinfo{year}{2001}).

\bibitem[{\citenamefont{Choi et~al.}(2009)\citenamefont{Choi, Louie, and
  Cohen}}]{Choi2009}
\bibinfo{author}{\bibfnamefont{H.~J.} \bibnamefont{Choi}},
  \bibinfo{author}{\bibfnamefont{S.~G.} \bibnamefont{Louie}}, \bibnamefont{and}
  \bibinfo{author}{\bibfnamefont{M.~L.} \bibnamefont{Cohen}},
  \bibinfo{journal}{Physical Review B} \textbf{\bibinfo{volume}{80}},
  \bibinfo{pages}{064503} (\bibinfo{year}{2009}).

\bibitem[{\citenamefont{Kolmogorov and Curtarolo}(2006)}]{Kolmogorov2006}
\bibinfo{author}{\bibfnamefont{A.~N.} \bibnamefont{Kolmogorov}}
  \bibnamefont{and}
  \bibinfo{author}{\bibfnamefont{S.}~\bibnamefont{Curtarolo}},
  \bibinfo{journal}{Phys. Rev. B} \textbf{\bibinfo{volume}{74}},
  \bibinfo{pages}{224507} (\bibinfo{year}{2006}).

\bibitem[{\citenamefont{Rosner et~al.}(2002)\citenamefont{Rosner,
  Kitaigorodsky, and Pickett}}]{Rosner2002}
\bibinfo{author}{\bibfnamefont{H.}~\bibnamefont{Rosner}},
  \bibinfo{author}{\bibfnamefont{A.}~\bibnamefont{Kitaigorodsky}},
  \bibnamefont{and} \bibinfo{author}{\bibfnamefont{W.~E.}
  \bibnamefont{Pickett}}, \bibinfo{journal}{Physical review letters}
  \textbf{\bibinfo{volume}{88}}, \bibinfo{pages}{127001}
  (\bibinfo{year}{2002}).

\bibitem[{\citenamefont{Calandra}(2007)}]{Calandra2007}
\bibinfo{author}{\bibfnamefont{M.}~\bibnamefont{Calandra}},
  \bibinfo{journal}{Physical review. C, Nuclear physics}
  \textbf{\bibinfo{volume}{75}}, \bibinfo{pages}{144506}
  (\bibinfo{year}{2007}), \bibinfo{note}{0556-2813}.

\bibitem[{\citenamefont{Lazicki et~al.}(2010)\citenamefont{Lazicki, Hemley,
  Pickett, and Yoo}}]{Lazicki2010}
\bibinfo{author}{\bibfnamefont{A.}~\bibnamefont{Lazicki}},
  \bibinfo{author}{\bibfnamefont{R.~J.} \bibnamefont{Hemley}},
  \bibinfo{author}{\bibfnamefont{W.~E.} \bibnamefont{Pickett}},
  \bibnamefont{and} \bibinfo{author}{\bibfnamefont{C.-S.} \bibnamefont{Yoo}},
  \bibinfo{journal}{Phys. Rev. B} \textbf{\bibinfo{volume}{82}},
  \bibinfo{pages}{180102} (\bibinfo{year}{2010}).

\bibitem[{\citenamefont{Bharathi et~al.}(2002)\citenamefont{Bharathi,
  Jemima~Balaselvi, Premila, Sairam, Reddy, Sundar, and
  Hariharan}}]{Bharathi2002}
\bibinfo{author}{\bibfnamefont{A.}~\bibnamefont{Bharathi}},
  \bibinfo{author}{\bibfnamefont{S.}~\bibnamefont{Jemima~Balaselvi}},
  \bibinfo{author}{\bibfnamefont{M.}~\bibnamefont{Premila}},
  \bibinfo{author}{\bibfnamefont{T.~N.} \bibnamefont{Sairam}},
  \bibinfo{author}{\bibfnamefont{G.~L.~N.} \bibnamefont{Reddy}},
  \bibinfo{author}{\bibfnamefont{C.~S.} \bibnamefont{Sundar}},
  \bibnamefont{and}
  \bibinfo{author}{\bibfnamefont{Y.}~\bibnamefont{Hariharan}},
  \bibinfo{journal}{Solid State Communications} \textbf{\bibinfo{volume}{124}},
  \bibinfo{pages}{423} (\bibinfo{year}{2002}).

\bibitem[{\citenamefont{Fujihisa et~al.}(2013)\citenamefont{Fujihisa, Nakamoto,
  Sakata, Shimizu, Matsuoka, Ohishi, Yamawaki, Takeya, and
  Gotoh}}]{Fujihisa2013}
\bibinfo{author}{\bibfnamefont{H.}~\bibnamefont{Fujihisa}},
  \bibinfo{author}{\bibfnamefont{Y.}~\bibnamefont{Nakamoto}},
  \bibinfo{author}{\bibfnamefont{M.}~\bibnamefont{Sakata}},
  \bibinfo{author}{\bibfnamefont{K.}~\bibnamefont{Shimizu}},
  \bibinfo{author}{\bibfnamefont{T.}~\bibnamefont{Matsuoka}},
  \bibinfo{author}{\bibfnamefont{Y.}~\bibnamefont{Ohishi}},
  \bibinfo{author}{\bibfnamefont{H.}~\bibnamefont{Yamawaki}},
  \bibinfo{author}{\bibfnamefont{S.}~\bibnamefont{Takeya}}, \bibnamefont{and}
  \bibinfo{author}{\bibfnamefont{Y.}~\bibnamefont{Gotoh}},
  \bibinfo{journal}{Phys. Rev. Lett.} \textbf{\bibinfo{volume}{110}},
  \bibinfo{pages}{235501} (\bibinfo{year}{2013}).

\bibitem[{\citenamefont{Oganov et~al.}(2010)\citenamefont{Oganov, Ma, Xu,
  Errea, Bergara, and Lyakhov}}]{Oganov2010}
\bibinfo{author}{\bibfnamefont{A.~R.} \bibnamefont{Oganov}},
  \bibinfo{author}{\bibfnamefont{Y.}~\bibnamefont{Ma}},
  \bibinfo{author}{\bibfnamefont{Y.}~\bibnamefont{Xu}},
  \bibinfo{author}{\bibfnamefont{I.}~\bibnamefont{Errea}},
  \bibinfo{author}{\bibfnamefont{A.}~\bibnamefont{Bergara}}, \bibnamefont{and}
  \bibinfo{author}{\bibfnamefont{A.~O.} \bibnamefont{Lyakhov}},
  \bibinfo{journal}{Proceedings of the National Academy of Sciences}
  \textbf{\bibinfo{volume}{107}}, \bibinfo{pages}{7646} (\bibinfo{year}{2010}).

\bibitem[{\citenamefont{Oganov et~al.}(2009)\citenamefont{Oganov, Chen, Gatti,
  Ma, Ma, Glass, Liu, Yu, Kurakevych, and Solozhenko}}]{Oganov2009}
\bibinfo{author}{\bibfnamefont{A.~R.} \bibnamefont{Oganov}},
  \bibinfo{author}{\bibfnamefont{J.}~\bibnamefont{Chen}},
  \bibinfo{author}{\bibfnamefont{C.}~\bibnamefont{Gatti}},
  \bibinfo{author}{\bibfnamefont{Y.}~\bibnamefont{Ma}},
  \bibinfo{author}{\bibfnamefont{Y.}~\bibnamefont{Ma}},
  \bibinfo{author}{\bibfnamefont{C.~W.} \bibnamefont{Glass}},
  \bibinfo{author}{\bibfnamefont{Z.}~\bibnamefont{Liu}},
  \bibinfo{author}{\bibfnamefont{T.}~\bibnamefont{Yu}},
  \bibinfo{author}{\bibfnamefont{O.~O.} \bibnamefont{Kurakevych}},
  \bibnamefont{and} \bibinfo{author}{\bibfnamefont{V.~L.}
  \bibnamefont{Solozhenko}}, \bibinfo{journal}{Nature}
  \textbf{\bibinfo{volume}{457}}, \bibinfo{pages}{863} (\bibinfo{year}{2009}).

\bibitem[{\citenamefont{Zarechnaya et~al.}(2008)\citenamefont{Zarechnaya,
  Dubrovinsky, Dubrovinskaia, Miyajima, Filinchuk, Chernyshov, and
  Dmitriev}}]{Zarechnaya2008}
\bibinfo{author}{\bibfnamefont{E.~Y.} \bibnamefont{Zarechnaya}},
  \bibinfo{author}{\bibfnamefont{L.}~\bibnamefont{Dubrovinsky}},
  \bibinfo{author}{\bibfnamefont{N.}~\bibnamefont{Dubrovinskaia}},
  \bibinfo{author}{\bibfnamefont{N.}~\bibnamefont{Miyajima}},
  \bibinfo{author}{\bibfnamefont{Y.}~\bibnamefont{Filinchuk}},
  \bibinfo{author}{\bibfnamefont{D.}~\bibnamefont{Chernyshov}},
  \bibnamefont{and} \bibinfo{author}{\bibfnamefont{V.}~\bibnamefont{Dmitriev}},
  \bibinfo{journal}{Science and Technology of Advanced Materials}
  \textbf{\bibinfo{volume}{9}}, \bibinfo{pages}{044209} (\bibinfo{year}{2008}).

\bibitem[{\citenamefont{Chepulskii and Curtarolo}(2009)}]{Chepulskii2009}
\bibinfo{author}{\bibfnamefont{R.~V.} \bibnamefont{Chepulskii}}
  \bibnamefont{and}
  \bibinfo{author}{\bibfnamefont{S.}~\bibnamefont{Curtarolo}},
  \bibinfo{journal}{Phys. Rev. B} \textbf{\bibinfo{volume}{79}},
  \bibinfo{pages}{134203} (\bibinfo{year}{2009}).

\bibitem[{\citenamefont{Yahia et~al.}(2008)\citenamefont{Yahia, Reckeweg,
  Gautier, Bauer, Schleid, Halet, and Saillard}}]{Yahia2008}
\bibinfo{author}{\bibfnamefont{M.~B.} \bibnamefont{Yahia}},
  \bibinfo{author}{\bibfnamefont{O.}~\bibnamefont{Reckeweg}},
  \bibinfo{author}{\bibfnamefont{R.}~\bibnamefont{Gautier}},
  \bibinfo{author}{\bibfnamefont{J.}~\bibnamefont{Bauer}},
  \bibinfo{author}{\bibfnamefont{T.}~\bibnamefont{Schleid}},
  \bibinfo{author}{\bibfnamefont{J.-F.} \bibnamefont{Halet}}, \bibnamefont{and}
  \bibinfo{author}{\bibfnamefont{J.-Y.} \bibnamefont{Saillard}},
  \bibinfo{journal}{Inorganic Chemistry} \textbf{\bibinfo{volume}{47}},
  \bibinfo{pages}{6137} (\bibinfo{year}{2008}).

\bibitem[{\citenamefont{Kolmogorov et~al.}(2012)\citenamefont{Kolmogorov, Shah,
  Margine, Kleppe, and Jephcoat}}]{Kolmogorov2012}
\bibinfo{author}{\bibfnamefont{A.~N.} \bibnamefont{Kolmogorov}},
  \bibinfo{author}{\bibfnamefont{S.}~\bibnamefont{Shah}},
  \bibinfo{author}{\bibfnamefont{E.~R.} \bibnamefont{Margine}},
  \bibinfo{author}{\bibfnamefont{A.~K.} \bibnamefont{Kleppe}},
  \bibnamefont{and} \bibinfo{author}{\bibfnamefont{A.~P.}
  \bibnamefont{Jephcoat}}, \bibinfo{journal}{Phys. Rev. Lett.}
  \textbf{\bibinfo{volume}{109}}, \bibinfo{pages}{075501}
  (\bibinfo{year}{2012}).

\bibitem[{\citenamefont{Albert and Schmitt}(1999)}]{Albert1999}
\bibinfo{author}{\bibfnamefont{B.}~\bibnamefont{Albert}} \bibnamefont{and}
  \bibinfo{author}{\bibfnamefont{K.}~\bibnamefont{Schmitt}},
  \bibinfo{journal}{Inorganic Chemistry} \textbf{\bibinfo{volume}{38}},
  \bibinfo{pages}{6159} (\bibinfo{year}{1999}).

\bibitem[{\citenamefont{Shah and Kolmogorov}()}]{supmat}
\bibinfo{author}{\bibfnamefont{S.}~\bibnamefont{Shah}} \bibnamefont{and}
  \bibinfo{author}{\bibfnamefont{A.}~\bibnamefont{Kolmogorov}},
  \emph{\bibinfo{title}{See supplemental information for this article with the
  publication in the physical review b.}}

\bibitem[{\citenamefont{Kolmogorov et~al.}(2010)\citenamefont{Kolmogorov, Shah,
  Margine, Bialon, Hammerschmidt, and Drautz}}]{Kolmogorov2010}
\bibinfo{author}{\bibfnamefont{A.~N.} \bibnamefont{Kolmogorov}},
  \bibinfo{author}{\bibfnamefont{S.}~\bibnamefont{Shah}},
  \bibinfo{author}{\bibfnamefont{E.~R.} \bibnamefont{Margine}},
  \bibinfo{author}{\bibfnamefont{A.~F.} \bibnamefont{Bialon}},
  \bibinfo{author}{\bibfnamefont{T.}~\bibnamefont{Hammerschmidt}},
  \bibnamefont{and} \bibinfo{author}{\bibfnamefont{R.}~\bibnamefont{Drautz}},
  \bibinfo{journal}{Physical review letters} \textbf{\bibinfo{volume}{105}},
  \bibinfo{pages}{217003} (\bibinfo{year}{2010}).

\bibitem[{\citenamefont{G.Bergerhoff and I.D.Brown}(1987)}]{ICSD}
\bibinfo{author}{\bibnamefont{G.Bergerhoff}} \bibnamefont{and}
  \bibinfo{author}{\bibnamefont{I.D.Brown}},
  \emph{\bibinfo{title}{Crystallographic Databases}}
  (\bibinfo{publisher}{International Union of Crystallography},
  \bibinfo{year}{1987}).

\bibitem[{\citenamefont{Kolmogorov}()}]{maise}
\bibinfo{author}{\bibfnamefont{A.}~\bibnamefont{Kolmogorov}},
  \urlprefix\url{www.maise-guide.org}.

\bibitem[{\citenamefont{Kresse and Furthm\"uller}(1996)}]{Kresse1996}
\bibinfo{author}{\bibfnamefont{G.}~\bibnamefont{Kresse}} \bibnamefont{and}
  \bibinfo{author}{\bibfnamefont{J.}~\bibnamefont{Furthm\"uller}},
  \bibinfo{journal}{Phys. Rev. B} \textbf{\bibinfo{volume}{54}},
  \bibinfo{pages}{11169} (\bibinfo{year}{1996}).

\bibitem[{\citenamefont{Bl\"{o}chl}(1994)}]{Blochl1994}
\bibinfo{author}{\bibfnamefont{P.~E.} \bibnamefont{Bl\"{o}chl}},
  \bibinfo{journal}{Physical Review B} \textbf{\bibinfo{volume}{50}},
  \bibinfo{pages}{17953} (\bibinfo{year}{1994}).

\bibitem[{\citenamefont{Monkhorst and Pack}(1976)}]{Pack1976}
\bibinfo{author}{\bibfnamefont{H.~J.} \bibnamefont{Monkhorst}}
  \bibnamefont{and} \bibinfo{author}{\bibfnamefont{J.~D.} \bibnamefont{Pack}},
  \bibinfo{journal}{Physical Review B} \textbf{\bibinfo{volume}{13}},
  \bibinfo{pages}{5188} (\bibinfo{year}{1976}).

\bibitem[{\citenamefont{Pack and Monkhorst}(1977)}]{Pack1977}
\bibinfo{author}{\bibfnamefont{J.~D.} \bibnamefont{Pack}} \bibnamefont{and}
  \bibinfo{author}{\bibfnamefont{H.~J.} \bibnamefont{Monkhorst}},
  \bibinfo{journal}{Physical Review B} \textbf{\bibinfo{volume}{16}},
  \bibinfo{pages}{1748} (\bibinfo{year}{1977}).

\bibitem[{\citenamefont{Perdew et~al.}(1996)\citenamefont{Perdew, Burke, and
  Ernzerhof}}]{Perdew1996}
\bibinfo{author}{\bibfnamefont{J.~P.} \bibnamefont{Perdew}},
  \bibinfo{author}{\bibfnamefont{K.}~\bibnamefont{Burke}}, \bibnamefont{and}
  \bibinfo{author}{\bibfnamefont{M.}~\bibnamefont{Ernzerhof}},
  \bibinfo{journal}{Physical Review Letters} \textbf{\bibinfo{volume}{77}},
  \bibinfo{pages}{3865} (\bibinfo{year}{1996}).

\bibitem[{\citenamefont{Alf\`{e}}(2009)}]{Alfe2009}
\bibinfo{author}{\bibfnamefont{D.}~\bibnamefont{Alf\`{e}}},
  \bibinfo{journal}{Computer Physics Communications}
  \textbf{\bibinfo{volume}{180}}, \bibinfo{pages}{2622} (\bibinfo{year}{2009}).

\bibitem[{\citenamefont{Rodgers and Villars}(1993)}]{Rodgers1993}
\bibinfo{author}{\bibfnamefont{J.~R.} \bibnamefont{Rodgers}} \bibnamefont{and}
  \bibinfo{author}{\bibfnamefont{P.}~\bibnamefont{Villars}},
  \bibinfo{journal}{Journal of Alloys and Compounds}
  \textbf{\bibinfo{volume}{197}}, \bibinfo{pages}{167} (\bibinfo{year}{1993}).

\bibitem[{\citenamefont{Paolo and et~al.}(2009)}]{Giannozzi2009}
\bibinfo{author}{\bibfnamefont{G.}~\bibnamefont{Paolo}} \bibnamefont{and}
  \bibinfo{author}{\bibnamefont{et~al.}}, \bibinfo{journal}{Journal of Physics:
  Condensed Matter} \textbf{\bibinfo{volume}{21}}, \bibinfo{pages}{395502}
  (\bibinfo{year}{2009}), \bibinfo{note}{0953-8984}.

\bibitem[{kq_()}]{kq_meshes}
\bibinfo{note}{We used the pseudopotentials Ca.pbe-nsp-van.UPF and
  B.pbe-n-van.UPF from www.quantum-espresso.org.}

\bibitem[{\citenamefont{Savrasov and Savrasov}(1996)}]{Savrasov1996}
\bibinfo{author}{\bibfnamefont{S.~Y.} \bibnamefont{Savrasov}} \bibnamefont{and}
  \bibinfo{author}{\bibfnamefont{D.~Y.} \bibnamefont{Savrasov}},
  \bibinfo{journal}{Phys. Rev. B} \textbf{\bibinfo{volume}{54}},
  \bibinfo{pages}{16487} (\bibinfo{year}{1996}).

\bibitem[{\citenamefont{Shannon}(1976)}]{Shannon1976}
\bibinfo{author}{\bibfnamefont{R.}~\bibnamefont{Shannon}},
  \bibinfo{journal}{Acta Crystallographica Section A}
  \textbf{\bibinfo{volume}{32}}, \bibinfo{pages}{751} (\bibinfo{year}{1976}).

\bibitem[{\citenamefont{Momma and Izumi}(2011)}]{Momma2011}
\bibinfo{author}{\bibfnamefont{K.}~\bibnamefont{Momma}} \bibnamefont{and}
  \bibinfo{author}{\bibfnamefont{F.}~\bibnamefont{Izumi}},
  \bibinfo{journal}{Journal of Applied Crystallography}
  \textbf{\bibinfo{volume}{44}}, \bibinfo{pages}{1272} (\bibinfo{year}{2011}).

\bibitem[{\citenamefont{Ji et~al.}(2011)\citenamefont{Ji, Zhang, Xu, and
  Zhao}}]{Ji2011}
\bibinfo{author}{\bibfnamefont{X.~H.} \bibnamefont{Ji}},
  \bibinfo{author}{\bibfnamefont{Q.~Y.} \bibnamefont{Zhang}},
  \bibinfo{author}{\bibfnamefont{J.~Q.} \bibnamefont{Xu}}, \bibnamefont{and}
  \bibinfo{author}{\bibfnamefont{Y.~M.} \bibnamefont{Zhao}},
  \bibinfo{journal}{Progress in Solid State Chemistry}
  \textbf{\bibinfo{volume}{39}}, \bibinfo{pages}{51} (\bibinfo{year}{2011}).

\bibitem[{\citenamefont{Young et~al.}(1999)\citenamefont{Young, Hall, Torelli,
  Fisk, Sarrao, Thompson, Ott, Oseroff, Goodrich, and Zysler}}]{Young1999}
\bibinfo{author}{\bibfnamefont{D.~P.} \bibnamefont{Young}},
  \bibinfo{author}{\bibfnamefont{D.}~\bibnamefont{Hall}},
  \bibinfo{author}{\bibfnamefont{M.~E.} \bibnamefont{Torelli}},
  \bibinfo{author}{\bibfnamefont{Z.}~\bibnamefont{Fisk}},
  \bibinfo{author}{\bibfnamefont{J.~L.} \bibnamefont{Sarrao}},
  \bibinfo{author}{\bibfnamefont{J.~D.} \bibnamefont{Thompson}},
  \bibinfo{author}{\bibfnamefont{H.~R.} \bibnamefont{Ott}},
  \bibinfo{author}{\bibfnamefont{S.~B.} \bibnamefont{Oseroff}},
  \bibinfo{author}{\bibfnamefont{R.~G.} \bibnamefont{Goodrich}},
  \bibnamefont{and} \bibinfo{author}{\bibfnamefont{R.}~\bibnamefont{Zysler}},
  \bibinfo{journal}{Nature} \textbf{\bibinfo{volume}{397}},
  \bibinfo{pages}{412} (\bibinfo{year}{1999}).

\bibitem[{\citenamefont{Rhyee and Cho}(2004)}]{Rhyee2004}
\bibinfo{author}{\bibfnamefont{J.-S.} \bibnamefont{Rhyee}} \bibnamefont{and}
  \bibinfo{author}{\bibfnamefont{B.~K.} \bibnamefont{Cho}},
  \bibinfo{journal}{Journal of Applied Physics} \textbf{\bibinfo{volume}{95}},
  \bibinfo{pages}{6675} (\bibinfo{year}{2004}).

\bibitem[{\citenamefont{Longuet-Higgins and De~V.~Roberts}(1954)}]{Longuet1954}
\bibinfo{author}{\bibfnamefont{H.~C.} \bibnamefont{Longuet-Higgins}}
  \bibnamefont{and}
  \bibinfo{author}{\bibfnamefont{M.}~\bibnamefont{De~V.~Roberts}},
  \bibinfo{journal}{Proceedings of the Royal Society of London. Series A.
  Mathematical and Physical Sciences} \textbf{\bibinfo{volume}{224}},
  \bibinfo{pages}{336} (\bibinfo{year}{1954}).

\bibitem[{\citenamefont{Li et~al.}(2011{\natexlab{a}})\citenamefont{Li, Yang,
  Li, Wang, Liang, and Gao}}]{Li2011a}
\bibinfo{author}{\bibfnamefont{M.}~\bibnamefont{Li}},
  \bibinfo{author}{\bibfnamefont{W.}~\bibnamefont{Yang}},
  \bibinfo{author}{\bibfnamefont{L.}~\bibnamefont{Li}},
  \bibinfo{author}{\bibfnamefont{H.}~\bibnamefont{Wang}},
  \bibinfo{author}{\bibfnamefont{S.}~\bibnamefont{Liang}}, \bibnamefont{and}
  \bibinfo{author}{\bibfnamefont{C.}~\bibnamefont{Gao}},
  \bibinfo{journal}{Physica B: Condensed Matter}
  \textbf{\bibinfo{volume}{406}}, \bibinfo{pages}{59}
  (\bibinfo{year}{2011}{\natexlab{a}}).

\bibitem[{\citenamefont{Li et~al.}(2011{\natexlab{b}})\citenamefont{Li, Yang,
  Cui, Hu, Liu, Tian, Liu, Han, and Gao}}]{Li2011b}
\bibinfo{author}{\bibfnamefont{Y.}~\bibnamefont{Li}},
  \bibinfo{author}{\bibfnamefont{J.}~\bibnamefont{Yang}},
  \bibinfo{author}{\bibfnamefont{X.}~\bibnamefont{Cui}},
  \bibinfo{author}{\bibfnamefont{T.}~\bibnamefont{Hu}},
  \bibinfo{author}{\bibfnamefont{C.}~\bibnamefont{Liu}},
  \bibinfo{author}{\bibfnamefont{Y.}~\bibnamefont{Tian}},
  \bibinfo{author}{\bibfnamefont{H.}~\bibnamefont{Liu}},
  \bibinfo{author}{\bibfnamefont{Y.}~\bibnamefont{Han}}, \bibnamefont{and}
  \bibinfo{author}{\bibfnamefont{C.}~\bibnamefont{Gao}},
  \bibinfo{journal}{Physica Status Solidi (b)} \textbf{\bibinfo{volume}{248}},
  \bibinfo{pages}{1162} (\bibinfo{year}{2011}{\natexlab{b}}).

\bibitem[{\citenamefont{Wei et~al.}(2011)\citenamefont{Wei, Yu, Li, Cheng, and
  Ji}}]{Wei2011}
\bibinfo{author}{\bibfnamefont{Y.-K.} \bibnamefont{Wei}},
  \bibinfo{author}{\bibfnamefont{J.-X.} \bibnamefont{Yu}},
  \bibinfo{author}{\bibfnamefont{Z.-G.} \bibnamefont{Li}},
  \bibinfo{author}{\bibfnamefont{Y.}~\bibnamefont{Cheng}}, \bibnamefont{and}
  \bibinfo{author}{\bibfnamefont{G.-F.} \bibnamefont{Ji}},
  \bibinfo{journal}{Physica B: Condensed Matter}
  \textbf{\bibinfo{volume}{406}}, \bibinfo{pages}{4476} (\bibinfo{year}{2011}).

\bibitem[{pba()}]{pban_LaB6}
\bibinfo{note}{This structure was proposed as a high pressure phase of LaB$_6$
  by Teredesai et. al. \cite{Teredesai2004}}.

\bibitem[{\citenamefont{Teredesai et~al.}(2004)\citenamefont{Teredesai, Muthu,
  Chandrabhas, Meenakshi, Vijayakumar, Modak, Rao, Godwal, Sikka, and
  Sood}}]{Teredesai2004}
\bibinfo{author}{\bibfnamefont{P.}~\bibnamefont{Teredesai}},
  \bibinfo{author}{\bibfnamefont{D.~V.~S.} \bibnamefont{Muthu}},
  \bibinfo{author}{\bibfnamefont{N.}~\bibnamefont{Chandrabhas}},
  \bibinfo{author}{\bibfnamefont{S.}~\bibnamefont{Meenakshi}},
  \bibinfo{author}{\bibfnamefont{V.}~\bibnamefont{Vijayakumar}},
  \bibinfo{author}{\bibfnamefont{P.}~\bibnamefont{Modak}},
  \bibinfo{author}{\bibfnamefont{R.~S.} \bibnamefont{Rao}},
  \bibinfo{author}{\bibfnamefont{B.~K.} \bibnamefont{Godwal}},
  \bibinfo{author}{\bibfnamefont{S.~K.} \bibnamefont{Sikka}}, \bibnamefont{and}
  \bibinfo{author}{\bibfnamefont{A.~K.} \bibnamefont{Sood}},
  \bibinfo{journal}{Solid State Communications} \textbf{\bibinfo{volume}{129}},
  \bibinfo{pages}{791} (\bibinfo{year}{2004}).

\bibitem[{\citenamefont{Godwal et~al.}(2009)\citenamefont{Godwal, Petruska,
  Speziale, Yan, Clark, Kruger, and Jeanloz}}]{Godwal2009}
\bibinfo{author}{\bibfnamefont{B.~K.} \bibnamefont{Godwal}},
  \bibinfo{author}{\bibfnamefont{E.~A.} \bibnamefont{Petruska}},
  \bibinfo{author}{\bibfnamefont{S.}~\bibnamefont{Speziale}},
  \bibinfo{author}{\bibfnamefont{J.}~\bibnamefont{Yan}},
  \bibinfo{author}{\bibfnamefont{S.~M.} \bibnamefont{Clark}},
  \bibinfo{author}{\bibfnamefont{M.~B.} \bibnamefont{Kruger}},
  \bibnamefont{and} \bibinfo{author}{\bibfnamefont{R.}~\bibnamefont{Jeanloz}},
  \bibinfo{journal}{Phys. Rev. B} \textbf{\bibinfo{volume}{80}},
  \bibinfo{pages}{172104} (\bibinfo{year}{2009}).

\bibitem[{\citenamefont{Xu et~al.}(2007)\citenamefont{Xu, Zhang, Cui, Li, Xie,
  Yu, Ma, and Zou}}]{Xu2007}
\bibinfo{author}{\bibfnamefont{Y.}~\bibnamefont{Xu}},
  \bibinfo{author}{\bibfnamefont{L.}~\bibnamefont{Zhang}},
  \bibinfo{author}{\bibfnamefont{T.}~\bibnamefont{Cui}},
  \bibinfo{author}{\bibfnamefont{Y.}~\bibnamefont{Li}},
  \bibinfo{author}{\bibfnamefont{Y.}~\bibnamefont{Xie}},
  \bibinfo{author}{\bibfnamefont{W.}~\bibnamefont{Yu}},
  \bibinfo{author}{\bibfnamefont{Y.}~\bibnamefont{Ma}}, \bibnamefont{and}
  \bibinfo{author}{\bibfnamefont{G.}~\bibnamefont{Zou}},
  \bibinfo{journal}{Phys. Rev. B} \textbf{\bibinfo{volume}{76}},
  \bibinfo{pages}{214103} (\bibinfo{year}{2007}).

\bibitem[{\citenamefont{Popov et~al.}(2012)\citenamefont{Popov, Baadji, and
  Sanvito}}]{Popov2012}
\bibinfo{author}{\bibfnamefont{I.}~\bibnamefont{Popov}},
  \bibinfo{author}{\bibfnamefont{N.}~\bibnamefont{Baadji}}, \bibnamefont{and}
  \bibinfo{author}{\bibfnamefont{S.}~\bibnamefont{Sanvito}},
  \bibinfo{journal}{Phys. Rev. Lett.} \textbf{\bibinfo{volume}{108}},
  \bibinfo{pages}{107205} (\bibinfo{year}{2012}).

\bibitem[{\citenamefont{Aroyo et~al.}(2006{\natexlab{a}})\citenamefont{Aroyo,
  Perez-Mato, Capillas, Kroumova, Ivantchev, Madariaga, Kirov, and
  Wondratschek}}]{Aroyo2006a}
\bibinfo{author}{\bibfnamefont{M.~I.} \bibnamefont{Aroyo}},
  \bibinfo{author}{\bibfnamefont{J.~M.} \bibnamefont{Perez-Mato}},
  \bibinfo{author}{\bibfnamefont{C.}~\bibnamefont{Capillas}},
  \bibinfo{author}{\bibfnamefont{E.}~\bibnamefont{Kroumova}},
  \bibinfo{author}{\bibfnamefont{S.}~\bibnamefont{Ivantchev}},
  \bibinfo{author}{\bibfnamefont{G.}~\bibnamefont{Madariaga}},
  \bibinfo{author}{\bibfnamefont{A.}~\bibnamefont{Kirov}}, \bibnamefont{and}
  \bibinfo{author}{\bibfnamefont{H.}~\bibnamefont{Wondratschek}},
  \bibinfo{journal}{Zeitschrift für Kristallographie}
  \textbf{\bibinfo{volume}{221}}, \bibinfo{pages}{15}
  (\bibinfo{year}{2006}{\natexlab{a}}).

\bibitem[{\citenamefont{Aroyo et~al.}(2006{\natexlab{b}})\citenamefont{Aroyo,
  Kirov, Capillas, Perez-Mato, and Wondratschek}}]{Ayoro2006_2}
\bibinfo{author}{\bibfnamefont{M.~I.} \bibnamefont{Aroyo}},
  \bibinfo{author}{\bibfnamefont{A.}~\bibnamefont{Kirov}},
  \bibinfo{author}{\bibfnamefont{C.}~\bibnamefont{Capillas}},
  \bibinfo{author}{\bibfnamefont{J.~M.} \bibnamefont{Perez-Mato}},
  \bibnamefont{and}
  \bibinfo{author}{\bibfnamefont{H.}~\bibnamefont{Wondratschek}},
  \bibinfo{journal}{Acta Crystallographica Section A}
  \textbf{\bibinfo{volume}{62}}, \bibinfo{pages}{115}
  (\bibinfo{year}{2006}{\natexlab{b}}).

\bibitem[{\citenamefont{Setyawan et~al.}(2011)\citenamefont{Setyawan, Gaume,
  Lam, Feigelson, and Curtarolo}}]{Setyawan2010}
\bibinfo{author}{\bibfnamefont{W.}~\bibnamefont{Setyawan}},
  \bibinfo{author}{\bibfnamefont{R.~M.} \bibnamefont{Gaume}},
  \bibinfo{author}{\bibfnamefont{S.}~\bibnamefont{Lam}},
  \bibinfo{author}{\bibfnamefont{R.~S.} \bibnamefont{Feigelson}},
  \bibnamefont{and}
  \bibinfo{author}{\bibfnamefont{S.}~\bibnamefont{Curtarolo}},
  \bibinfo{journal}{ACS Combinatorial Science} \textbf{\bibinfo{volume}{13}},
  \bibinfo{pages}{382} (\bibinfo{year}{2011}).

\bibitem[{\citenamefont{Lee and Wang}(2005)}]{Lee2005}
\bibinfo{author}{\bibfnamefont{B.}~\bibnamefont{Lee}} \bibnamefont{and}
  \bibinfo{author}{\bibfnamefont{L.-W.} \bibnamefont{Wang}},
  \bibinfo{journal}{Applied Physics Letters} \textbf{\bibinfo{volume}{87}},
  \bibinfo{pages}{262509} (\bibinfo{year}{2005}).

\bibitem[{\citenamefont{Zalkin and Templeton}(1953)}]{Zalkin1953}
\bibinfo{author}{\bibfnamefont{A.}~\bibnamefont{Zalkin}} \bibnamefont{and}
  \bibinfo{author}{\bibfnamefont{D.~H.} \bibnamefont{Templeton}},
  \bibinfo{journal}{Acta Crystallographica} \textbf{\bibinfo{volume}{6}},
  \bibinfo{pages}{269} (\bibinfo{year}{1953}).

\bibitem[{\citenamefont{Lafferty}(1951)}]{Lafferty1950}
\bibinfo{author}{\bibfnamefont{J.~M.} \bibnamefont{Lafferty}},
  \bibinfo{journal}{Journal of Applied Physics} \textbf{\bibinfo{volume}{22}},
  \bibinfo{pages}{299} (\bibinfo{year}{1951}).

\bibitem[{\citenamefont{Yin and Pickett}(2008)}]{Yin2008}
\bibinfo{author}{\bibfnamefont{Z.~P.} \bibnamefont{Yin}} \bibnamefont{and}
  \bibinfo{author}{\bibfnamefont{W.~E.} \bibnamefont{Pickett}},
  \bibinfo{journal}{Phys. Rev. B} \textbf{\bibinfo{volume}{77}},
  \bibinfo{pages}{035135} (\bibinfo{year}{2008}).

\bibitem[{\citenamefont{Johnson and Daane}(1961)}]{Johnson1961}
\bibinfo{author}{\bibfnamefont{R.~W.} \bibnamefont{Johnson}} \bibnamefont{and}
  \bibinfo{author}{\bibfnamefont{A.~H.} \bibnamefont{Daane}},
  \bibinfo{journal}{The Journal of Physical Chemistry}
  \textbf{\bibinfo{volume}{65}}, \bibinfo{pages}{909} (\bibinfo{year}{1961}).

\bibitem[{\citenamefont{Schmitt et~al.}(2006)\citenamefont{Schmitt,
  Blaschkowski, Eichele, and Meyer}}]{Schmitt2006}
\bibinfo{author}{\bibfnamefont{R.}~\bibnamefont{Schmitt}},
  \bibinfo{author}{\bibfnamefont{B.}~\bibnamefont{Blaschkowski}},
  \bibinfo{author}{\bibfnamefont{K.}~\bibnamefont{Eichele}}, \bibnamefont{and}
  \bibinfo{author}{\bibfnamefont{H.~J.} \bibnamefont{Meyer}},
  \bibinfo{journal}{Inorganic Chemistry} \textbf{\bibinfo{volume}{45}},
  \bibinfo{pages}{3067} (\bibinfo{year}{2006}).

\bibitem[{\citenamefont{Liu}(2010)}]{Liu2010}
\bibinfo{author}{\bibfnamefont{Z.}~\bibnamefont{Liu}}, \bibinfo{journal}{Appl.
  Phys. Lett.} \textbf{\bibinfo{volume}{96}}, \bibinfo{pages}{031903}
  (\bibinfo{year}{2010}).

\bibitem[{\citenamefont{Mutterities}(1967)}]{Mutterities1967}
\bibinfo{author}{\bibfnamefont{E.~L.} \bibnamefont{Mutterities}},
  \emph{\bibinfo{title}{The Chemistry of Boron and its Compounds}}
  (\bibinfo{publisher}{John Wiley $\&$ Sons, Inc.}, \bibinfo{year}{1967}).

\bibitem[{\citenamefont{Burdett and Canadell}(1991)}]{Burdett1991}
\bibinfo{author}{\bibnamefont{Burdett}} \bibnamefont{and}
  \bibinfo{author}{\bibnamefont{Canadell}}, \bibinfo{journal}{ChemInform}
  \textbf{\bibinfo{volume}{22}}, \bibinfo{pages}{7207} (\bibinfo{year}{1991}).

\bibitem[{\citenamefont{Bialon et~al.}(2011)\citenamefont{Bialon,
  Hammerschmidt, Drautz, Shah, Margine, and Kolmogorov}}]{bialon2011}
\bibinfo{author}{\bibfnamefont{A.~F.} \bibnamefont{Bialon}},
  \bibinfo{author}{\bibfnamefont{T.}~\bibnamefont{Hammerschmidt}},
  \bibinfo{author}{\bibfnamefont{R.}~\bibnamefont{Drautz}},
  \bibinfo{author}{\bibfnamefont{S.}~\bibnamefont{Shah}},
  \bibinfo{author}{\bibfnamefont{E.~R.} \bibnamefont{Margine}},
  \bibnamefont{and} \bibinfo{author}{\bibfnamefont{A.~N.}
  \bibnamefont{Kolmogorov}}, \bibinfo{journal}{Applied Physics Letters}
  \textbf{\bibinfo{volume}{98}}, \bibinfo{eid}{081901}
  (pages~\bibinfo{numpages}{3}) (\bibinfo{year}{2011}).

\bibitem[{\citenamefont{Andersson and Lundstrom}(1968)}]{Andersson1968}
\bibinfo{author}{\bibfnamefont{S.}~\bibnamefont{Andersson}} \bibnamefont{and}
  \bibinfo{author}{\bibfnamefont{T.}~\bibnamefont{Lundstrom}},
  \bibinfo{journal}{Acta Chem Scand.} \textbf{\bibinfo{volume}{22}},
  \bibinfo{pages}{3103} (\bibinfo{year}{1968}).

\bibitem[{\citenamefont{Niu et~al.}(2012)\citenamefont{Niu, Wang, Chen, Li, Li,
  Lazar, Podloucky, and Kolmogorov}}]{Niu2012}
\bibinfo{author}{\bibfnamefont{H.}~\bibnamefont{Niu}},
  \bibinfo{author}{\bibfnamefont{J.}~\bibnamefont{Wang}},
  \bibinfo{author}{\bibfnamefont{X.-Q.} \bibnamefont{Chen}},
  \bibinfo{author}{\bibfnamefont{D.}~\bibnamefont{Li}},
  \bibinfo{author}{\bibfnamefont{Y.}~\bibnamefont{Li}},
  \bibinfo{author}{\bibfnamefont{P.}~\bibnamefont{Lazar}},
  \bibinfo{author}{\bibfnamefont{R.}~\bibnamefont{Podloucky}},
  \bibnamefont{and} \bibinfo{author}{\bibfnamefont{A.~N.}
  \bibnamefont{Kolmogorov}}, \bibinfo{journal}{Phys. Rev. B}
  \textbf{\bibinfo{volume}{85}}, \bibinfo{pages}{144116}
  (\bibinfo{year}{2012}).

\bibitem[{\citenamefont{Naslain et~al.}(1973)\citenamefont{Naslain, Guette, and
  Barret}}]{Naslain1973}
\bibinfo{author}{\bibfnamefont{M.~R.} \bibnamefont{Naslain}},
  \bibinfo{author}{\bibfnamefont{A.}~\bibnamefont{Guette}}, \bibnamefont{and}
  \bibinfo{author}{\bibfnamefont{M.}~\bibnamefont{Barret}},
  \bibinfo{journal}{Journal of Solid State Chemistry}
  \textbf{\bibinfo{volume}{8}}, \bibinfo{pages}{68 } (\bibinfo{year}{1973}),
  ISSN \bibinfo{issn}{0022-4596}.

\bibitem[{\citenamefont{Tang et~al.}(2009)\citenamefont{Tang, Sanville, and
  Henkelman}}]{Tang2009}
\bibinfo{author}{\bibfnamefont{W.}~\bibnamefont{Tang}},
  \bibinfo{author}{\bibfnamefont{E.}~\bibnamefont{Sanville}}, \bibnamefont{and}
  \bibinfo{author}{\bibfnamefont{G.}~\bibnamefont{Henkelman}},
  \bibinfo{journal}{Journal of Physics: Condensed Matter}
  \textbf{\bibinfo{volume}{21}}, \bibinfo{pages}{084204}
  (\bibinfo{year}{2009}).

\bibitem[{\citenamefont{Zhang et~al.}(2012)\citenamefont{Zhang, Legut, Lin,
  Zhao, Mao, and Veprek}}]{Zhang2012}
\bibinfo{author}{\bibfnamefont{R.~F.} \bibnamefont{Zhang}},
  \bibinfo{author}{\bibfnamefont{D.}~\bibnamefont{Legut}},
  \bibinfo{author}{\bibfnamefont{Z.~J.} \bibnamefont{Lin}},
  \bibinfo{author}{\bibfnamefont{Y.~S.} \bibnamefont{Zhao}},
  \bibinfo{author}{\bibfnamefont{H.~K.} \bibnamefont{Mao}}, \bibnamefont{and}
  \bibinfo{author}{\bibfnamefont{S.}~\bibnamefont{Veprek}},
  \bibinfo{journal}{Phys. Rev. Lett.} \textbf{\bibinfo{volume}{108}},
  \bibinfo{pages}{255502} (\bibinfo{year}{2012}).

\bibitem[{\citenamefont{Buzea and Yamashita}(2001)}]{Buzea2001}
\bibinfo{author}{\bibfnamefont{C.}~\bibnamefont{Buzea}} \bibnamefont{and}
  \bibinfo{author}{\bibfnamefont{T.}~\bibnamefont{Yamashita}},
  \bibinfo{journal}{Superconductor Science and Technology}
  \textbf{\bibinfo{volume}{14}}, \bibinfo{pages}{R115} (\bibinfo{year}{2001}).

\bibitem[{\citenamefont{Hermann et~al.}(2012)\citenamefont{Hermann, Ashcroft,
  and Hoffmann}}]{Hermann2012}
\bibinfo{author}{\bibfnamefont{A.}~\bibnamefont{Hermann}},
  \bibinfo{author}{\bibfnamefont{N.~W.} \bibnamefont{Ashcroft}},
  \bibnamefont{and} \bibinfo{author}{\bibfnamefont{R.}~\bibnamefont{Hoffmann}},
  \bibinfo{journal}{Inorganic Chemistry} \textbf{\bibinfo{volume}{51}},
  \bibinfo{pages}{9066} (\bibinfo{year}{2012}).

\bibitem[{\citenamefont{Sands et~al.}(1961)\citenamefont{Sands, Cline, Zalkin,
  and Hoenig}}]{Sands1961}
\bibinfo{author}{\bibfnamefont{D.~E.} \bibnamefont{Sands}},
  \bibinfo{author}{\bibfnamefont{C.~F.} \bibnamefont{Cline}},
  \bibinfo{author}{\bibfnamefont{A.}~\bibnamefont{Zalkin}}, \bibnamefont{and}
  \bibinfo{author}{\bibfnamefont{C.~L.} \bibnamefont{Hoenig}},
  \bibinfo{journal}{Acta Crystallographica} \textbf{\bibinfo{volume}{14}},
  \bibinfo{pages}{309} (\bibinfo{year}{1961}).

\bibitem[{\citenamefont{Vajeeston et~al.}(2012)\citenamefont{Vajeeston,
  Ravindran, and Fjellvag}}]{Vajeeston2012}
\bibinfo{author}{\bibfnamefont{P.}~\bibnamefont{Vajeeston}},
  \bibinfo{author}{\bibfnamefont{P.}~\bibnamefont{Ravindran}},
  \bibnamefont{and} \bibinfo{author}{\bibfnamefont{H.}~\bibnamefont{Fjellvag}},
  \bibinfo{journal}{RSC Adv.} \textbf{\bibinfo{volume}{2}},
  \bibinfo{pages}{11687} (\bibinfo{year}{2012}).

\bibitem[{\citenamefont{Goncharov and Struzhkin}(2003)}]{Goncharov2003}
\bibinfo{author}{\bibfnamefont{A.~F.} \bibnamefont{Goncharov}}
  \bibnamefont{and} \bibinfo{author}{\bibfnamefont{V.~V.}
  \bibnamefont{Struzhkin}}, \bibinfo{journal}{Physica C: Superconductivity}
  \textbf{\bibinfo{volume}{385}}, \bibinfo{pages}{117 } (\bibinfo{year}{2003}).

\bibitem[{\citenamefont{Etourneau and Hagenmuller}(1985)}]{Etourneau1985}
\bibinfo{author}{\bibfnamefont{J.}~\bibnamefont{Etourneau}} \bibnamefont{and}
  \bibinfo{author}{\bibfnamefont{P.}~\bibnamefont{Hagenmuller}},
  \bibinfo{journal}{Philosophical Magazine Part B}
  \textbf{\bibinfo{volume}{52}}, \bibinfo{pages}{589} (\bibinfo{year}{1985}).

\bibitem[{\citenamefont{Gauzzi et~al.}(2007)\citenamefont{Gauzzi, Takashima,
  Takeshita, Terakura, Takagi, Emery, H\'erold, Lagrange, and
  Loupias}}]{Gauzzi2007}
\bibinfo{author}{\bibfnamefont{A.}~\bibnamefont{Gauzzi}},
  \bibinfo{author}{\bibfnamefont{S.}~\bibnamefont{Takashima}},
  \bibinfo{author}{\bibfnamefont{N.}~\bibnamefont{Takeshita}},
  \bibinfo{author}{\bibfnamefont{C.}~\bibnamefont{Terakura}},
  \bibinfo{author}{\bibfnamefont{H.}~\bibnamefont{Takagi}},
  \bibinfo{author}{\bibfnamefont{N.}~\bibnamefont{Emery}},
  \bibinfo{author}{\bibfnamefont{C.}~\bibnamefont{H\'erold}},
  \bibinfo{author}{\bibfnamefont{P.}~\bibnamefont{Lagrange}}, \bibnamefont{and}
  \bibinfo{author}{\bibfnamefont{G.}~\bibnamefont{Loupias}},
  \bibinfo{journal}{Phys. Rev. Lett.} \textbf{\bibinfo{volume}{98}},
  \bibinfo{pages}{067002} (\bibinfo{year}{2007}).

\bibitem[{\citenamefont{Xing-Qiu et~al.}(2010)\citenamefont{Xing-Qiu, Fu, and
  Franchini}}]{Chen2010}
\bibinfo{author}{\bibfnamefont{C.}~\bibnamefont{Xing-Qiu}},
  \bibinfo{author}{\bibfnamefont{C.~L.} \bibnamefont{Fu}}, \bibnamefont{and}
  \bibinfo{author}{\bibfnamefont{C.}~\bibnamefont{Franchini}},
  \bibinfo{journal}{Journal of Physics: Condensed Matter}
  \textbf{\bibinfo{volume}{22}}, \bibinfo{pages}{292201}
  (\bibinfo{year}{2010}).

\end{thebibliography}

\end{document}